\documentclass[10pt, reqno]{amsart}
\usepackage{amsmath, amssymb, amsfonts, amstext, verbatim, amsthm, mathtools}

\input xy
\xyoption{all}
\usepackage{setspace}
\usepackage{enumerate}
\usepackage{prettyref}

\newrefformat{prop}{Proposition~\ref{#1}}
\newrefformat{cor}{Corollary~\ref{#1}}
\newrefformat{thm}{Theorem~\ref{#1}}
\newrefformat{lem}{Lemma~\ref{#1}}
\newrefformat{def}{Definition~\ref{#1}}

\theoremstyle{plain}
\newtheorem{thm}{Theorem}[section]
\newtheorem{lem}[thm]{Lemma}

\newtheorem{conj}[thm]{Conjecture}

\newtheorem{quest}[thm]{Question}

\theoremstyle{definition}
\newtheorem{rem}[thm]{Remark}
\newtheorem{defn}[thm]{Definition}
\newtheorem{ex}[thm]{Example}

\newcommand{\op}[1]{\!\!\mathop{\rm ~#1}\nolimits}

\renewcommand{\tilde}{\widetilde}

\begin{document}

\title{Curved String topology and Tangential Fukaya Categories II}
\author{Daniel Pomerleano}
\maketitle
 
\section{Introduction} 
   In this thesis, we study deformations of dg-categories to smooth and proper dg-categories. Our interest in the smooth and proper condition comes from the following theorem of Kontsevich and Soibelman \cite{KS} that \vskip 10 pt \begin{em} Given a compact and smooth $\mathbb{Z}/2\mathbb{Z}$ graded Calabi-Yau $A_{\infty}$ algebra $B$ for which the Hodge to De-Rham spectral sequence degenerates, a choice of splitting for this spectral sequence gives rise to a Cohomological Field Theory (CFT). \end{em} \vskip 10 pt

In section 2 of this thesis, we will recall these notions and results in more detail. The most basic example of such a category is the (dg-enhanced) derived category of quasicoherent sheaves $QCoh(\mathcal{X})$ on a compact and smooth Calabi-Yau variety. This category satisfies all of the above conditons, and the resulting field theory is known as the \emph{B-model} for this Calabi-Yau variety. Homological Mirror Symmetry \cite{Kontsevich} predicts that the associated CFT is expected to be equivalent to the Gromov-Witten CFT on the mirror CY variety $\mathcal{X}^{\vee}$. \vskip 10 pt
  
   Now consider $\mathcal{Y}$ to be a smooth but non-compact Calabi-Yau variety. In this setting, $QCoh(\mathcal{Y})$ is a non-compact Calabi-Yau category, and by a modified version of the theorem of Kontsevich and Soibelman, we can get a so-called \begin{em}positive-output\end{em} topological quantum field theory (TQFT). The Landau-Ginzburg model uses deformation theory to compactify these theories by deforming the above category by a superpotential $w$, which is an algebraic function with a proper critical set. Recent work \cite{Preygel,Lin-Pomerleano} shows that this gives rise to a TQFT. For more on curved algebras, see section 3. \vskip 10 pt 
   
  Section 2 of the present paper recalls that a similar situation occurs in topology. Namely, there is a positive output TQFT called \begin{em} string topology \end{em} for a compact oriented manifold $\mathcal{Q}$ associated to the dg-category of dg-modules  $mod(C_*(\Omega \mathcal{Q}))$ over the dg algebra $C_*(\Omega \mathcal{Q})$ \cite{Lurie}, where $\Omega \mathcal{Q}$ denotes the based loop space of $\mathcal{Q}$ at some arbitrary point, $pt$. Throughout this thesis, all coefficients are taken to be $\mathbb{C}$, the field of complex numbers. As we explain below, this category is a smooth but not compact category. The relationship with string topology is revealed by the following calculation for the Hochschild homology:
  
 $$ \mathbf{HH}_*(C_*(\Omega \mathcal{Q}))\cong C_*(\mathcal{LQ})$$

  There is also a natural compact Calabi-Yau category associated to such a manifold, the category of modules over $C^*(\mathcal{Q})$, which however is not smooth. Such categories give rise to TQFT's with \begin{em} positive-input\end{em}. When $\mathcal{Q}$ is simply connected, these two algebras are related via Koszul duality. Namely, the inclusion $pt \to Q $, induces a module structure:
  
   $$ C^*(\mathcal{Q}) \to \mathbb{C} $$
   
 The vector space $\mathbb{C}$ can also be thought of as a module over $C_*(\Omega \mathcal{Q})$ by regarding it as the trivial local system. As discussed in \cite{BluCohTel}, the following isomorphisms hold: \vskip 10 pt

$$\mathbb{R}Hom_{C^*(\mathcal{Q})}(\mathbb{C},\mathbb{C})\cong C_*(\Omega \mathcal{Q})$$  $$C^*(\mathcal{Q})\cong \mathbb{R}Hom_{C_*(\Omega \mathcal{Q})}(\mathbb{C},\mathbb{C})$$

 and in fact this gives rise to fully faithful functors:
 $$perf(C_*(\Omega\mathcal{Q}) \to mod(C^*(\mathcal{Q}))$$ and $$perf(C^*(\mathcal{Q})^{op}) \to mod(C_*(\Omega \mathcal{Q})^{op})$$

Here $perf(C_*(\Omega\mathcal{Q})$ or $perf(C^*(\mathcal{Q})^{op})$ denotes the subcategory of perfect modules, which is defined for the reader below. Nevertheless, $\mathbb{C}$ is not a compact generator in the category $mod(C_*(\Omega \mathcal{Q}))$ which means that Koszul duality does not give rise to an equivalence of the full derived categories. Following \cite{Abouzaid}, Section 1.1 also reviews the relationship between string topology and the Fukaya category of $T^*\mathcal{Q}$, which provides a geometric way of thinking about this Koszul duality.\vskip 10 pt

 The case where $\mathcal{Q}$ is $T^n= S^1\times S^1\times \cdots \times S^1$, served as motivation for the present work. Dyckerhoff \cite{Dyckerhoff} proved the following theorem: 

\begin{thm} Let $w$ be a function on $\mathbb{C}[[x_1,x_2,\ldots,x_n]]$ with isolated singularities. The object $\mathbb{C}$ is a compact generator for $MF(\mathbb{C}[[x_1,x_2,\ldots,x_n]],w)$. Otherwise stated, $Hom_{MF(\mathbb{C}[[x_1,x_2,\ldots,x_n]],w)}(\mathbb{C},-)$ defines an equivalence of categories: $$ MF(\mathbb{C}[[x_1,x_2,\ldots,x_n]],w) \to mod(Hom_{MF(\mathbb{C}[[x_1,x_2,\ldots,x_n]],w)}(\mathbb{C},\mathbb{C})) $$ \end{thm} 

Here $MF(\mathbb{C}[[x_1,x_2,\ldots,x_n]],w)$ denotes the category of matrix factorizations, whose definition occupies much of section 3. The relationship between this theorem and the previous discussion is that $C_*(\Omega T^n)$ is isomorphic to $\mathbb{C}[z_1,z_1^{-1},z_2,z_2^{-1},\ldots z_n,z_n^{-1}]$, the Laurent polynomial ring in several variables. As $T^n= S^1\times S^1\ldots \times S^1$ is not simply connected, we complete at the augmentation ideal of this ring to obtain $\mathbb{C}[[x_1,x_2,\ldots,x_n]]$. In such cases, $MF(\mathbb{C}[[x_1,x_2,\ldots,x_n]],w)$ defines a quantum field theory. This result can be viewed as a deformed Koszul duality in the sense that $Hom_{MF(\mathbb{C}[[x_1,x_2,\ldots,x_n]],w)}(\mathbb{C},\mathbb{C}) \cong H^*(T^n)$ with a deformed $A_{\infty}$ structure :

      $$ m_{\ell}: H^*(T^n)^{\otimes \ell} \to H^*(T^n) $$
      
     the coefficients of which can be derived from $w$ in a direct manner \cite{Dyckerhoff, Efimov}. \vskip 10 pt
  
   In this thesis, we will consider simply connected manifolds $\mathcal{Q}$ whose minimal models are pure Sullivan algebras, which are generalizations of complete intersection rings (see section 3 for the precise definition). Section 4 of our paper makes precise and then gives an answer to the following question:
 
\begin{quest} If $C^*(\mathcal{Q})$ is a pure Sullivan algebra and given an element $w\in Z(C_*(\Omega \mathcal{Q}))$, when is $\mathbb{C}$ a compact generator of $MF(C_*(\Omega \mathcal{Q}),w)$ defining an equivalence with $mod(H^*(C^*(\mathcal{Q})),m_\ell)$? \end{quest} 
 
 In section 5, we make the Hochschild cohomology of $MF(C_*(\Omega \mathcal{Q}),w)$ explicit, prove that the deformed category is still Calabi-Yau and deduce the degeneration of the aforementioned Hodge-de Rham spectral sequence. We will examine our condition in the special case that the differential of our pure Sullivan algebra is quadratic. In section 6, we give some comments on the pure Sullivan condition.  \vskip 10 pt
 
 As mentioned earlier, morally, one can think of a potential $w$ as ``compactifying" the field theory. In section 7 of our paper, inspired by a program of \cite{Seidel}, we explain how the simplest of our theories, such as when $\mathcal{Q}=\mathbb{C}P^n$ or $S^n$, arises by geometrically compactifying the cotangent bundles $T^*\mathbb{C}P^n$ and $T^*S^n$ inside of a certain root stack. The definition of the root stack is explained at the beginning of section 7. A slightly more precise statement of our result is that we realize our category $MF(C_*(\Omega \mathcal{Q}),w)$ as being equivalent to the subcategory of the Fukaya category of the root stack generated by the zero section. We also discuss how to place these constructions fit into constructions in Symplectic Field Theory. \vskip 10 pt

 Section 8 discusses the above compactifications from the point of view of Homological Mirror Symmetry and SYZ fibrations, which allows us to refine the equivalence in the previous section for some low dimensional examples. \vskip 10 pt

Section 9 rounds out this thesis by using SYZ fibrations to generalize the above results to $A_n$ plumbings of cotangent bundles of $S^2$. We begin by proving homological mirror symmetry for these plumbings. Then we move on to studying the curved deformations of their wrapped Fukaya categories. \vskip 10 pt

The author would like to thank his adviser Constantin Teleman for sharing his vision of mathematics for many years and for suggesting the general direction of this research. Denis Auroux taught the author about mirror symmetry and symplectic geometry which have played an unexpectedly large role in this work. He also read an early draft of this work and made many suggestions to improve the exposition. The author would also like to thank Mohammed Abouzaid and Yakov Eliashberg for inspiring discussions about symplectic geometry and Anatoly Preygel for teaching him about curved dg-algebras.

\section{Background and Algebraic setup}

All gradings referred to below will follow homological grading conventions. Given a pre-triangulated dg-category $\mathcal{C}$, we denote its associated triangulated category by $[\mathcal{C}]$. For invariants derived from these categories, such as $\mathbf{HH}^*(\mathcal{C})$ or $\mathbf{HH}_*(\mathcal{C})$, bold font will be used when it is important that the construction be carried out at the chain level. Any use of functors such as $Hom$ or $\otimes$ is always assumed to be derived. \vskip 10 pt

Recall that a dg-module (or $A_{\infty}$-module) $N$ over a dg-algebra (or $A_{\infty}$-algebra) $\mathcal{A}$  is \begin{em}perfect\end{em} if it is contained in the smallest idempotent-closed triangulated subcategory of $[mod(\mathcal{A})]$ generated by $\mathcal{A}$. In general, unless explicitly stated otherwise, we will use the term ``generation" to mean what symplectic geometers usually call ``split- generation''. That is to say, given a collection of objects $\mathbf{O}_i$ in a triangulated category $\mathcal{C}$, the subcategory generated by $\mathbf{O}_i$ will be the smallest idempotent-closed triangulated subcategory containing $\mathbf{O}_i$.  

\begin{defn} A dg-algebra $\mathcal{A}$ over $\mathbb{C}$ is \begin{em} compact \end{em} if $\mathcal{A}$ is perfect as a $\mathbb{C}$ module (in this special case this simply says that $\mathcal{A}$ is equivalent to a finite dimensional vector space). A dg-algebra $\mathcal{A}$ is \begin{em}smooth\end{em} if $\mathcal{A}$ is perfect as an $\mathcal{A}-\mathcal{A}$ bimodule.\end{defn} 

A very useful criterion for smoothness is given by the notion of finite-type of To\"en
 and Vaquie \cite{Toen2}. 
 
 \begin{defn} A dg-algebra $\mathcal{A}$ is of finite type if it is a homotopy retract in the homotopy category of dg-algebras of a free algebra $(\mathbb{C}\langle v_1,v_2,\ldots,v_n\rangle,d)$ with  $dv_j \in \mathbb{C}\langle v_1,v_2,\ldots,v_{j-1}\rangle$ \end{defn}
 
 \begin{lem} If $\mathcal{A}$ is of finite type then $\mathcal{A}$ is smooth. The converse is also true if $\mathcal{A}$ is assumed to be compact. \end{lem}
 
 \begin{lem} With the notation of the previous section, the dg-algebra $C_*(\Omega \mathcal{Q})$ is smooth. \end{lem}
 
In the simply connected case, this follows from the classical Adams-Hilton construction \cite{Adams} and the above theorem of To\"en-Vaquie. Consider a cellular model for $\mathcal{Q}$ with cells in dimension $\leq dim(\mathcal{Q})$ and no 1-cells. Let $\mathcal{A}$ denote the tensor algebra generated by variables $e^b_i$, $deg(e^b_i)= b-1$, for all $n$ and $i$, where $e^b_i$ are in bijection with cells of dimension $b$ in the cell decomposition. \vskip 10 pt

Let $A_b$ denote the algebra generated by cells of dimension $\leq b$. For a given cell, $d(e^b_i)=z$, where $z$ is defined as the pushforward the canonical class in $H_{b-2}(\Omega S^{b-1})$ under the attaching map $f:S^{b-1} \to \mathcal{Q}$. Thus, we can see that the differential $d$ maps $d: A_b \to A_{b-1}$ and that the algebra is of finite type. The theorem remains true in the non-simply connected case, but the proof is more complicated \cite{Kontsevich II}. \vskip 10 pt 

\begin{defn} A dg-algebra $\mathcal{A}$ is Calabi-Yau of dimension $n$ if $Hom_{\mathbb{C}}(\mathcal{A},\mathbb{C}) \cong \mathcal{A}[-n]$ as $\mathcal{A}$-$\mathcal{A}$ bimodules.
\end{defn}

\begin{ex} Given a manifold $\mathcal{Q}$, the algebra $C^*(\mathcal{Q})$ is Calabi-Yau by Poincare duality. \end{ex}

\begin{defn} A smooth dg-algebra $\mathcal{A}$ is non-compact Calabi-Yau if 

$$Hom_{(\mathcal{A}^e)^{op}}(\mathcal{A},\mathcal{A}^e) \cong \mathcal{A}[n]$$ 

as $\mathcal{A}$-$\mathcal{A}$ bimodules.
\end{defn}

\begin{ex} Given a manifold $\mathcal{Q}$, we can again let $\mathcal{A}= C_*(\Omega\mathcal{Q})$. This is a non-campact Calabi-Yau. To see this, note that by smoothness we have that:
 
   $$ Hom_{\mathcal{A}^e}(Hom_{(\mathcal{A}^e)^{op}}(\mathcal{A},\mathcal{A}^e),\mathcal{A}[n]) \cong \mathcal A \otimes_{\mathcal{A}^e} \mathcal{A}[n]$$  
As noted above we have that:

$$\mathcal A \otimes_{\mathcal{A}^e} \mathcal{A}[n] \cong C_*(\mathcal{L}\mathcal{Q})[n]$$

The fundamental class of $\mathcal{Q}$ provides the desired isomorphism 

$$Hom_{(\mathcal{A}^e)^{op}}(\mathcal{A},\mathcal{A}^e) \cong \mathcal{A}[n]$$ 

\end{ex}

Next we recall a tiny bit about how the duality between $C^*(\mathcal{Q})$ and $C_*(\Omega \mathcal{Q})$ for compact simply connected manifolds is reflected in the symplectic geometry of their cotangent bundles. Consider $T^*\mathcal{Q}$ with its standard symplectic form $d\theta$. Let $b$, the background class, be the class in $H^*(T^*\mathcal{Q},\mathbb{Z}_2)$ given by the pullback of the second Stieffel-Whitney class of $\mathcal{Q}$. The classical Fukaya category, $Fuk(T^*\mathcal{Q},b)$, consists of (twisted complexes of) compact exact Lagrangian submanifolds $\mathcal{L}$ such that the restriction of $b$ to $\mathcal{L}$ is $w_2(\mathcal{L})$. For any two Lagrangian submanifolds $\mathcal{L}_1$ and $\mathcal{L}_2$, their morphisms are defined by the Floer homology groups with coefficients in $\mathbb{Q}$, $HF^*(\mathcal{L}_1,\mathcal{L}_2)$ \cite{Seidel II}. The zero section $\mathcal{Q}$ defines such an object. We have the following description of its endomorphisms:

   $$ HF^*(\mathcal{Q},\mathcal{Q}) \cong C^*(\mathcal{Q}) $$

 For Liouville symplectic manifolds such as $T^*\mathcal{Q}$, it is convenient to consider a version of the Fukaya category, known as the \begin{em} wrapped Fukaya category \end{em}, $WFuk(T^*\mathcal{Q})$, which allows us to incorporate non-compact Lagrangians into the Fukaya category. An important example of such an object is the cotangent fibre to a point $q$, denoted as $T_q$. For its definition see \cite{Abouzaid}. One very important property of the wrapped Fukaya category is that we have a natural fully faithful functor: 

$$i: Fuk(T^*\mathcal{Q}) \to WFuk(T^*\mathcal{Q}) $$ 

Abouzaid proves the following theorem: \vskip 10 pt

\begin{thm} The cotangent fibre strongly generates the wrapped Fukaya category of $T^*\mathcal{Q}$ with background class $b \in H^*(T^*\mathcal{Q},\mathbb{Z}_2)$ given by the pullback of the second Stieffel-Whitney class of $\mathcal{Q}$. The triangulated closure of the wrapped Fukaya category is equivalent to the category $perf(C_*(\Omega \mathcal{Q}))$. \end{thm}

The second sentence follows from the first because of the following description of the endomorphisms of the cotangent fibre:

$$WHF^*(T_q,T_q) \cong C_*(\Omega \mathcal{Q})$$

\section{Pure Sullivan algebras and Curved algebras} 

We consider Pure Sullivan dg-algebras $\mathcal{B}$ of the form: 
 $$ (\wedge V, d)=(\mathbb{C}[x_1,...x_n]\otimes \bigwedge(\beta_1,...\beta_m), d(\beta_i)=f_i(x_1,\ldots, x_n),d(x_j)=0) $$ 
where the $deg(x_i)$ are even and negative, the functions $f_i$ have no linear term, and the $deg(\beta_i)$ are odd $>1$. 
We further assume that $dim(H^*(\mathcal{B}))<\infty$. \vskip 10 pt

One of the underlying ideas of Chevalley-Eilenberg theory is that such algebras determine an $L_{\infty}$ model $\mathfrak{g}$ for $\mathcal{B}$. We can define an algebra $\mathcal{A}$ which is the universal enveloping algebra $U\mathfrak{g}$ of these Lie algebras. We briefly explain certain ideas from rational homotopy theory which will be used extensively below. For more details, the reader is encouraged to consult \cite{Felix}. To a simply connected space of finite type, $\mathcal{M}$, one can assign an $L_{\infty}$ algebra $\mathfrak{g}$ = $\pi_*(\Omega(\mathcal{M}))\otimes \mathbb{Q}$, with Whitehead-Samelson bracket. To recover $C^*(\mathcal{M})$, one considers the Chevalley-Eilenberg complex $C^*(\mathfrak{g})$, a canonical complex that computes Lie-algebra cohomology with coefficients in the trivial module. We have a quasi-isomorphism,

    $$ C^*(\mathfrak{g}) \to C^*(\mathcal{M})  $$ 

Furthermore, one can show that $U\mathfrak{g}$ is then quasi-isomorphic to $C_*(\Omega(\mathcal{M}))$. To go in the other direction, the theory of rational homotopy allows us to assign a space, $\mathcal{M}$, well defined up to rational homotopy equivalence, to a 1-connected dga or equivalently a connected dg-Lie algebras of finite type. \vskip 10 pt

In the case of pure Sullivan algebras $\mathcal{B}$, there is a concrete description of the universal enveloping $A_{\infty}$ algebra $\mathcal{A}$. Using the homological perturbation lemma, we have an explicit $A_{\infty}$ model for $\mathcal{A}$ of the form $$(Sym(\mathfrak{g}_{even}) \otimes \Lambda(\mathfrak{g}_{odd}),m_n)$$

 A formula for the higher multiplications appears in section 3 of \cite{Baranovsky}. For our purposes, we note the following facts. First, the strict morphism of the abelian Lie algebra $\pi_{even}(\Omega(Q)) \to \mathfrak{g}$ corresponds to the inclusion of $Sym(\mathfrak{g}_{even})\cong \mathbb{C}[u_1,\ldots,u_m] \to \mathcal{A}$. The higher multiplications $m_n$ are multi-linear in these variables for $n \geq 3$. Finally, we have that the $A_{\infty}$ algebra is strictly unital and the augmentation $U\mathfrak{g} \to \mathbb{C}$ defined by killing $\mathfrak{g}U\mathfrak{g}$ is also a strict morphism. \vskip 10 pt
 
 The reader should be warned that in the presence of quadratic terms in the $f_i$, the above identification with $Sym(\mathfrak{g}_{even}) \otimes \Lambda(\mathfrak{g}_{odd})$ is only an identification of vector spaces. In other words, there can be a non-trivial Lie bracket $B:\mathfrak{g}_{odd}\otimes \mathfrak{g}_{odd} \to \mathfrak{g}_{even}$, which means that forgetting higher products, $U\mathfrak{g}$ is a Clifford algebra over $Sym(\mathfrak{g}_{even})$.  It also seems worth pointing out that the even variables $u_i$ can be thought of as being Koszul dual to the odd variables $\beta_i$. Meanwhile the variables in $\mathfrak{g}_{odd}$, from here on denoted as $e_j$, are dual to the even variables $x_j$ above.  \vskip 10 pt
 
Next, we discuss how to define an appropriate category of matrix factorizations. This section adopts the ideas of the foundational work \cite{Preygel} to our non-commutative context. For concreteness, let us consider as before the above $A_{\infty}$-algebra $\mathcal{A}$, and an element $w \in \mathbb{C}[u_1,\ldots, u_m]$ of degree $2j-2$. For example, if $M=\mathbb{C}P^n$, we have the following specific model: $$U\mathfrak{g}= \mathbb{C}[u] \otimes \Lambda(e), m_{n+1}(e,e,e,\ldots,e)=u$$ We can then consider potentials of the form $w=u^d$. \vskip 10 pt

We define a variable $x$ of degree $2j-2$. The element $w$ defines a mapping from $$w:\mathbb{C}[x] \to \mathcal{A}$$ and we can consider the $A_{\infty}$ algebra $\mathcal{A}_0=(\mathcal{A}[e], de=w)$, where $e$ now has degree $2j-1$. 

\begin{defn} We define $Pre(MF(\mathcal{A},w))$, to be the full subcategory of $mod(\mathcal{A}_0)$ consisting of modules which are perfect over $\mathcal{A}$. \end{defn}

This category is equipped with a natural $\mathbb{C}[[t]]$ (degree $t=-2j$) linear structure which we will now describe. 
\begin{rem} Of course, $\mathbb{C}[[t]]$ as a graded ring is usually denoted $\mathbb{C}[t]$. The notation $\mathbb{C}[[t]]$ is simply to note that it should be treated as a topological ring. For example, given a graded vector space $V$, 

$$V[[t]]_n = \prod_{k \geq 0} V_{m+2jk}$$

 This will be distinct from $V[t]$ if $V$ is not homologically bounded from above. \end{rem}

We begin with the description of the $\mathbb{C}[[t]]$-linear structure from an abstract point of view and then give more concrete descriptions. We observe that:

$$Pre(MF(\mathcal{A},w))\cong RHom_{\mathbb{C}[x]}(Perf(\mathbb{C}),Perf(\mathcal{A}))$$   

the category of colimit preserving functors\cite{Toen}. We describe this construction a bit more below. For the reader who is become disoriented with the notation, notice that the $\mathbb{C}[x]$ structure on the right-hand side comes from the above algebra map $w$.
 
This category of functors is acted upon the category $RHom_{\mathbb{C}[x]}(Perf(\mathbb{C}),Perf(\mathbb{C})$ by convolution. Let $\alpha$ denote a variable of degree $2j-1$. Then there is an isomorphism:

$$RHom_{\mathbb{C}[x]}(Perf(\mathbb{C}),Perf(\mathbb{C})\cong D_{fin}(\mathbb{C}[\alpha]/\alpha^2) $$ 

Here $D_{fin}(\mathbb{C}[\alpha]/\alpha^2)$ denotes the subcategory category of modules over $\mathbb{C}[\alpha]/\alpha^2$ which are homologically finite over $\mathbb{C}$. Next we notice that Koszul duality provides an equivalence : 

$$D_{fin}(\mathbb{C}[\alpha]/\alpha^2) \cong Perf(\mathbb{C}[[t]])$$ 

The aforementioned $\mathbb{C}[[t]]$(degree $t=-2n$) linear structure now arises in view of the natural equivalence between (idempotent complete, pre-triangulated) module categories over $Perf(\mathbb{C}[[t]])$ and ordinary $\mathbb{C}[[t]]$-linear, (idempotent complete, pre-triangulated) dg categories.\vskip 10 pt

This description of the action of $t$ is relatively obscure and so we now aim to unravel it and make it more concrete. The object $\mathbb{C}$ in $D_{fin}(\mathbb{C}[\alpha]/\alpha^2)$ acts via the identity. The action for the module $\mathbb{C}[\alpha]/\alpha^2$ can be described by considering the composition of the two adjoint-functors:
 
 $$i_*:RHom_{\mathbb{C}[x]}(Perf(\mathbb{C}),Perf(\mathcal{A}))\to RHom_{\mathbb{C}[x]}(Perf(\mathbb{C}[x]),Perf(\mathcal{A})) $$
 
 $$ i^*: RHom_{\mathbb{C}[x]}(Perf(\mathbb{C}[x]),Perf(\mathcal{A})) \to RHom_{\mathbb{C}[x]}(Perf(\mathbb{C}),Perf(\mathcal{A}))$$

In view of the fact that $i^*\circ i_*(N) \cong \mathbb{C}\otimes_{\mathbb{C}[x]} N $, and the fact that N and $\mathbb{C}$ are perfect over $\mathcal{A}$ and $\mathbb{C}[x]$ respectively, it follows that $i^*\circ i_*(N)$ is perfect over $\mathcal{A}_0$.

One can resolve the module $\mathbb{C}$ over $\mathbb{C}[\alpha]/\alpha^2$ by the standard Koszul resolution:

    $$ \mathbb{C} \cong  \bigoplus_k \frac{\mathbb{C}[\alpha]}{\alpha^2}[u^k/k!], du=\alpha $$

Applying this resolution to an object in M in $Pre(MF(\mathcal{A},w))$, we conclude that:

$$ Hom_{Pre(MF(\mathcal{A},w))}(M,N)= (Hom_{Perf(\mathcal{A})}(M,N)[[t]],d) $$

where 

$$ d: \phi \to d_{\mathcal{A}}(\phi) + t(\phi \circ e \wedge+ e \wedge \circ \phi ))$$

The differential $d_{\mathcal{A}}$ denotes the differential on $Hom_{Perf(\mathcal{A})}(M,N)$. In this equation $t$ acts in the natural way. \vskip 10 pt

The above construction generalizes the construction of the category of singularities for ordinary commutative rings \cite{Orlov}. It is natural to ask how this $\mathbb{C}[[t]]$ linear structure arises from deformation theory or how it can be expressed in a way that resembles the usual category of matrix factorizations. The first step is to define a reasonable category of modules for the topologically complete unital, augmented, curved $A_{\infty}$ algebra 

$$(\mathcal{A}[[t]],tw)$$ 

The following construction was outlined in \cite{Pomerleano}. We denote by $\mathcal{A}_{+}$ the quotient $\mathcal{A}/\mathbb{C}$. We note that the element $tw$ also defines a Maurer-Cartan element in $\mathbf{HH}^*(\mathcal{A},\mathcal{A})[[t]]$.  Such a Maurer-Cartan solution allows us to twist the differential on  
  
  $$(\bigoplus_n \mathcal{A}_{+}^{\otimes n}[[t]],d_A) $$ 
  
  by the differential determined by the formula:

\[td_w: a_0 \otimes a_1 \otimes \cdots \otimes a_n \mapsto \sum_{i=0}^{n-1} (-1)^{i+1} t a_0 \otimes a_1 \otimes \cdots \otimes a_i \otimes W \otimes a_{i+1} \otimes \cdots \otimes a_n. \]

giving rise to a topologically complete coalgebra: 

$$\mathbf{C}=(\bigoplus \mathcal{A}_{+}^{\otimes n}[[t]],d_{\mathcal{A}}+td_w)$$. 

We can look now at modules over this coalgebra which are topologically free over $\mathbb{C}[[t]]$, topologically cofree as modules over the underlying coalgebra, and are perfect over $A$ when $t=0$. We denote this category by $comod(\mathbf{C})$.
 
 \begin{lem} The functor $\mathcal{F}: M\to ((\bigoplus_n \mathcal{A}_{+}^{\otimes n} \otimes M)[[t]],d_{M/\mathcal{A}}+te\wedge)$ defines a fully faithful functor:
 $$ Pre(MF(\mathcal{A},w)) \to comod(\mathbf{C})$$ \end{lem} 

Ignoring differentials for ease of notation we have that:

$$ Hom(\mathcal{F}(M),\mathcal{F}(N))= Hom^{top}_{\mathbb{C}[[t]]}( (\bigoplus_n \mathcal{A}_{+}^{\otimes n}\otimes M)[[t]], N[[t]])$$

We can identify this with:

$$ Hom_{\mathbb{C}}(\bigoplus_n \mathcal{A}_{+}^{\otimes n} \otimes M, N)[[t]] $$

The differential on this complex is again the differential $d_{\mathcal{A}} + t(\phi \circ e \wedge+ e \wedge \circ \phi )$. 

Finally, we define $$MF(\mathcal{A},w) = Pre(MF(\mathcal{A},w))\otimes_{\mathbb{C}[[t]]}\mathbb{C}((t))  \cong  comod(\mathbf{C})\otimes_{\mathbb{C}[[t]]}\mathbb{C}((t)) $$ 

The fact that $i^*\circ i_*(N)$ is perfect for any object in $Pre(MF(\mathcal{A},w))$ implies just as in the usual case that 

  $$[MF(\mathcal{A},w)] \cong [Pre(MF(\mathcal{A},w))]/[Perf(\mathcal{A}_0)]$$

It is often convenient to work with the formal Ind-completion $Ind(MF(\mathcal{A},w))$ which we shall denote by $MF^{\infty}(\mathcal{A},w)$. \vskip 10 pt

   We have constructed a category of curved modules for a curved $A_{\infty}$ algebra which arises as a deformation of an uncurved $A_{\infty}$ algebra.
A recent article by Positselski \cite{Positselski, PositselskiII} explains a similar construction of a module category for curved $A_{\infty}$ algebras $\mathcal{A}$ over complete local rings $k$ where the potential is contained inside $m\mathcal{A}$. This is consistent with the above philosophy that it is necessary to treat $\mathbb{C}[[t]]$ as a topological ring to construct a good category of modules. There is no good general construction of a module category over a curved $A_{\infty}$ algebra. \vskip 10 pt

To illustrate this point, it might be useful to consider the case of ordinary matrix factorizations, namely pairs $(R,w)$ where $R$ is a commutative ring as above. If one considers it as a two-periodic curved $A_{\infty}$-algebra, various authors \cite{CaldararuTu} have noted that the category of comodules over the two-periodic bar algebra:
 
$$ \bigoplus_n R^{\otimes n},d_R + d_w $$
 
is always zero. Thus it is important to consider the bar complex over $\mathbb{C}[[t]]$, calculate the corresponding Hom-sets and then invert $t$. As an example, consider the case when the function $w$ has isolated singularities. 
The naive Hochshild complex $$ (\prod Hom_{\mathbb{C}}(R^{n+1},R), d_{Hoch} + [w,])$$ is always zero.
However we have a quasi-isomorphism as two periodic complexes: 

$$ (\prod Hom_{\mathbb{C}}(R^{n+1},R)((t)),d_{Hoch} +t[w,]) \cong  (\bigoplus Hom_{\mathbb{C}}(R^{n+1},R), d_{Hoch} + [w,]) $$   

This latter complex computes the Jacobian ring as one would expect.  

\section{ The criterion for generation}

In this section we discuss a criterion for smoothness and properness of the category $MF(\mathcal{A},w)$. To state the criterion, we must consider the category of curved bimodules $$MF(\mathcal{A}\otimes \mathcal{A}^{op},w\otimes1-1\otimes w)$$ and we define $\mathbf{HH}^*(MF(\mathcal{A},w))$ to be $Hom_{MF(\mathcal{A}\otimes \mathcal{A}^{op},w\otimes1-1\otimes w)}(\mathcal{A},\mathcal{A})$. Using either description of our category, this can be computed explicitly as: 

$$\mathbf{HH}^*(MF(\mathcal{A},w))  \cong (\mathbf{HH}^*(\mathcal{A},\mathcal{A})((t)),d_{\mathcal{A}}+[tw,]) $$

The following is the analogue of Dyckerhoff's theorem for our situation:

\begin{thm} If $\mathbf{HH}^*(MF(\mathcal{A},w))$ is finite over $\mathbb{C}((t))$, then $\mathbb{C}((t))$ generates the category $MF(\mathcal{A},w)$.\end{thm} 
We have an action of $\mathbb{C}[u_1,\ldots,u_m]$ on $DSing(\mathcal{A}_0)$ which factors through the complex 
$\mathbf{HH}^*(MF(\mathcal{A},w))$. For any u in $\mathbb{C}[u_1,\ldots,u_m]$, we let $K_u$ be the diagram
\[
\xymatrix {\mathbb{C}[u_1,\ldots,u_m] \ar[r]^u & \mathbb{C}[u_1,\ldots,u_m] }
\]

Finally, for the sequence $\bar{u}=(u_1,\ldots,u_m)$ we define $$ K_{\bar{u}^p} = \otimes K_{u_i^p} $$.  
 
With this notation in hand, we consider the colimit of the diagram:

$$ K_{\bar{u}} \to K_{\bar{u}^2}\to K_{\bar{u}^3} \ldots $$

which we denote by $R\Gamma_m$. For any object $\mathbf{O}$ in $MF(\mathcal{A},w)$,  we have an augmentation 

$$R\Gamma_m \otimes_ {\mathbb{C}[u_1,\ldots,u_m]}\mathbf{O} \to \mathbf{O} \to cone(e)$$

Because the action of $\mathbb{C}[u_1,\ldots,u_m]$ factors as above, we see that such an m-equivalence is in fact an equivalence and can conclude that $cone(e)$ is zero. \vskip 10 pt

Now the objects $K_{\bar{u}^i}\otimes \mathbf{O}$ are in the triangulated subcategory generated by $\mathbb{C}$ because $\mathcal{A}$ is finitely generated as a module over $\mathbb{C}[u_1,\ldots,u_m]$. The objects $\mathbf{O}$ are compact in $MF^{\infty}(\mathcal{A},w)$ and can be expressed as a colimit of $K_{\bar{u}^i} \otimes \mathbf{O}$. Therefore we can conclude that $\mathbf{O}$ is a direct summand of one of the $K_{\bar{u}^i}(\mathbf{D})\otimes \mathbf{O}$ generated by $\mathbb{C}$ as well. 

\begin{rem} The same argument goes through if the ideal of the kernel of the above ring homomorphism is $I$. Namely, in this situation, the category is generated by $\mathbb{C}[u_1,\ldots u_m]/I \otimes_{\mathbb{C}[u_1,\ldots u_m]}\mathcal{A} $. We isolate the case where $\mathbb{C}[u_1,\ldots u_m]/I$ is finite dimensional because it has the most relevance to topological field theories. \end{rem}

To discuss homological smoothness, we must consider the  category: 
             
             $$RHom^c_{\mathbb{C}((t))}(MF^{\infty}(\mathcal{A},w),MF^{\infty}(\mathcal{A},w))$$ 

the category of continuous endofunctors in the sense of \cite{Toen}, which we now describe. Given two dg-categories $\mathcal{C}_1$ and $\mathcal{C}_2$, the naive category of dg-functors Hom($\mathcal{C}_1$,$\mathcal{C}_2$) is not well behaved with respect to quasi-equivalence of dg-categories. \vskip 10 pt

 To\"en \cite{Toen} proved that there is a model structure on the category of dg-categories where weak equivalences are given by quasi-equivalences and the category $RHom(\mathcal{C}_1,\mathcal{C}_2)$ is a derived functor with respect to this model structure. We have a natural inclusion $RHom^c(\mathcal{C}_1,\mathcal{C}_2) \subset RHom(\mathcal{C}_1,\mathcal{C}_2)$ of all functors which commute with arbitrary colimits. \vskip 10 pt

To\"en proves that for if a co-complete dg-category $\mathcal{C}$ has a compact generator $\mathbf{O}$, and is thus equivalent to the category of modules $mod(Hom(\mathbf{O},\mathbf{O})^{op})$, then we have that \cite{Toen}: 

$$ RHom^c(\mathcal{C},\mathcal{C}) \cong mod(Hom(\mathbf{O},\mathbf{O})\otimes Hom(\mathbf{O},\mathbf{O})^{op}) $$

Thus we have the following theorem:

\begin{thm} $ RHom^c_{\mathbb{C}((t))}(MF^{\infty}(\mathcal{A},w) ,MF^{\infty}(\mathcal{A},w)) \cong MF^{\infty}(\mathcal{A}\otimes \mathcal{A}^{op}, w\otimes 1- 1\otimes w)$ \end{thm}

This follows because $\mathbb{C}$ also generates the category $MF^{\infty}(\mathcal{A}\otimes \mathcal{A}^{op}, w\otimes 1- 1\otimes w)$. Let $t_1$ be the deformation parameter corresponding to $Pre(MF(\mathcal{A},w))$ and $t_2$ be the deformation parameter corresponding to $Pre(MF(\mathcal{A}^{op},-w))$. We know that $Hom_{\mathcal{A}}(\mathbb{C},\mathbb{C})$ is homologically bounded above. This implies that the mapping:

$$Hom_{A}(\mathbb{C},\mathbb{C})[[t_1]]\otimes_{\mathbb{C}[[t]]} Hom_{\mathcal{A}}(\mathbb{C},\mathbb{C})[[t_2]] \to Hom_{\mathcal{A}\otimes \mathcal{A}^{op}}(\mathbb{C},\mathbb{C})[[t_1,t_2]]\otimes\mathbb{C}[[t]]$$  

is an equivalence. We achieve the result by inverting $t$.

\begin{rem} One of the downsides of the formalism that we have chosen is that the above theorem will fail almost uniformly if the generating object $\mathbf{O}$ in $MF^{\infty}(A,w)$ does not have the property that $Hom(\mathbf{O},\mathbf{O})$ is homologically bounded from above, essentially because the above mapping will almost never be an equivalence. This situation might be improved by giving a definition of the monoidal category of dg-categories where one makes use of completed tensor products. \end{rem}

\section{The Calabi-Yau property, Hochschild cohomology and the Degeneration Conjecture}

Next we will show that the Calabi-Yau condition for our category $MF(\mathcal{A},w)$ follows from $\mathcal{A}$ being non-compact Calabi-Yau. We first briefly recall why $\mathcal{A}$ is non-compact Calabi-Yau. Recall that this means that there is an isomorphism of $\mathcal{A}-\mathcal{A}$- bimodules 

$$ RHom_{\mathcal{A}^e}(\mathcal{A},\mathcal{A}^e) \cong \mathcal{A}[n] $$ 

The dg-algebra $\mathcal{B}$ is rationally elliptic. By results in Chapter [35] of the book \cite{Felix} the algebra $H^*(\mathcal{B})$ is a Poincare duality algebra. Now results in \cite{Costello} prove that the deformation theory of $C_{\infty}$ algebras and Frobenius $C_{\infty}$ algebras with a fixed trace coincide. By applying the perturbation lemma and viewing the dg-algebra $\mathcal{B}$ as a deformation of $H^*(\mathcal{B})$, this implies that the Frobenius structure on $H^*(\mathcal{B})$ enhances naturally to a Calabi-Yau structure on $\mathcal{B}$. \vskip 10 pt

 Next we have the following theorem proved in \cite{Van Den Bergh}:
 
\begin{thm} Let $\mathcal{A}$ be a homologically smooth algebra concentrated in degree $\geq 0$. Then a cyclic $A_{\infty}$ structure on its Koszul dual algebra gives rise to a non-compact Calabi-Yau structure on $\mathcal{A}$. \end{thm}

Now that $\mathcal{A}$ is seen to be Calabi-Yau, we show that this implies the property for $MF(\mathcal{A},w)$.
To prove the Calabi-Yau property for $MF(\mathcal{A},w)$, we note that we have a relative dualizing functor:

$$ D: MF(\mathcal{A},w) \to MF(\mathcal{A}^{op},-w)^{op} $$
$$ M \to RHom_{\mathcal{A}}(M,\mathcal{A}) $$

$D$ is manifestly an equivalence. Note that for any object $O$ in $MF(\mathcal{A}\otimes \mathcal{A}^{op},w\otimes1-1\otimes w)$ we have that 

$$ Hom_{MF(\mathcal{A}\otimes \mathcal{A}^{op},w\otimes1-1\otimes w)}(\mathbb{C},O) \cong Hom_{MF(\mathcal{A}\otimes \mathcal{A}^{op},w\otimes1-1\otimes w)}(D(O), D(\mathbb{C}))$$ 

We know that $$D(\mathbb{C}) \cong \mathbb{C}[n]$$
      
      and because $\mathcal{A}$ is non-compact Calabi-Yau, we have that  $$D(\Delta) \cong \Delta[n] $$
      
From here we learn that for the diagonal $$Hom_{MF}(\Delta, \mathbb{C}) \cong Hom_{MF}(\mathbb{C},\Delta)$$   
  
 Next we denote the complex $Hom_{MF(\mathcal{A},w)}(\mathbb{C},\mathbb{C})$ by  $\mathcal{D}$. 
This means that there is an isomorphism of $\mathcal{D}-\mathcal{D}$- bimodules 

$$ RHom_{\mathcal{D}^e}(\mathcal{D},\mathcal{D}^e) \cong \mathcal{D}[n] $$

which implies the Calabi-Yau condition for $MF(\mathcal{A},w)$. \vskip 10 pt

We can make our condition on finiteness of $\mathbf{HH}^*(MF(\mathcal{A},w))$ more tractable by considering the deformation theory of the pure Sullivan algebra $\mathcal{B}$ itself. As noted in the introduction, for any simply connected space of finite type, we have fully faithful functors induced by the $C^*(\mathcal{Q})-C_*(\Omega \mathcal{Q})$ bimodule $\mathbb{C}$. It then follows from a result of Keller \cite{Keller} that for such a fully faithful functor there is a canonical equivalence in the homotopy category of $B(\infty)$ algebras:
 
 $$\mathbf{HH}^*(C^*(\mathcal{Q}),C^*(\mathcal{Q}))\cong \mathbf{HH}^*(C_*(\Omega \mathcal{Q}),C_*(\Omega \mathcal{Q}))$$
 
 In particular these two Koszul dual algebras have equivalent formal deformation theories. Suppose that, more generally, we consider a commutative algebra free- graded commutative model $(\bigwedge V,d)$ where $V$ is a finite dimensional graded vector space. There is a very explicit complex quasi-isomorphic as a dg-Lie algebra to $\mathbf{HH}^*((\bigwedge V,d),(\bigwedge V,d))$. Recall that $T^{poly}(V)$ is the Lie-algebra of polyvector fields on $\bigwedge V$ with Schouten bracket. Part of Kontsevich's formality theorem says that the HKR map: $$T^{poly}(V) \to \mathbf{HH}^*(\bigwedge V)$$ is the first Taylor coefficient in an $L_{\infty}$ quasi-isomorphism between the two. \vskip 10 pt
 
 We can think of the derivation $d$ as corresponding to a vector-field $v$. It follows from a spectral sequence argument that the HKR map gives a quasi-isomorphism:

$$(T^{poly}(V),[v,-]) \to \mathbf{HH}^*((\bigwedge V,d),(\bigwedge V,d))$$

\begin{lem} This map can be corrected to an $L_{\infty}$ quasi-isomorphism. In the case of a pure Sullivan algebra, the first Taylor coefficient agrees with the HKR map.\end{lem}

Denote by $f_n$ the Taylor coefficients of the Kontsevich formality morphism. We consider a modified $L_{\infty}$ map $$\widetilde{f_n}:(T^{poly}(V),[v,-]) \to \mathbf{HH}^*(\bigwedge(V,d))$$ given by the formula 

$$\widetilde{f_n}(x_1,\cdots,x_n)=\sum 1/k! f_{k+n}(v,v,\cdots,v,x_1,x_2, \cdots,x_n)$$

This sum, seemingly a sum consisting of infinitely many terms, makes sense in this case for the following reason: for all $\gamma_1,\ldots \gamma_m$ in $T^{poly}(V)$,

 $$ f_{n+k}(v,v,\cdots,v,x_1, x_2,\cdots,x_n)(\gamma_1,\ldots \gamma_m) \text{ vanishes unless } 2(n+k)+m-2=k+\sum(|x_i|) $$ 
 
It is easy to check that this defines an $L_{\infty}$ map between the two complexes in question. We must further convince ourselves that the map $\widetilde{f_1}$ remains a quasi-isomorphism. By the above equation, we know that for any $x \in T^{poly}(V) $ such that $f_1(x) \in HH^m$, we have that $$\widetilde{f_1}(x)=f_1(x)+\alpha$$ where $\alpha \in HH^{<m}$ Assuming the map is surjective onto $HH^{<m}$, we learn by induction that the map is surjective onto $HH^{\leq m}$ as well. Injectivity of the map on homology is also clear. \vskip 10 pt

In general the formula for our map $\widetilde{f_1}$ can be computed explicitly from work of \cite{Calaque} but is very complicated, the coefficients of the map being given in terms of Bernoulli numbers. In the case of interest, we actually know more, namely, we have the following:

\begin{lem} If the dg algebra is pure Sullivan the map $\widetilde{f_1}$ agrees with the map $f_1$. \end{lem}

 Following [Ca], we denote our coordinates by $u_k$, and write and $v=\sum v_i \partial_i$. There is a matrix valued one form given by
$$\Gamma_{ji}=\sum_{k} \partial_i \partial_k v_j du_k$$
and define
$$\theta=\sum_{n>0}c_n i_{tr}(\Gamma^n)$$
where the $c_n$ are certain rational coefficients.

Calaque proves that $$\widetilde{f_1}=f_1(e^\theta)$$

Now if $v=\sum f(x_1,�,x_n)d/de_i$ as above, then the matrix is strictly upper triangular, the trace of any power of it is therefore zero, and thus $\widetilde{f_1}=f_1$. \vskip 10 pt 
 
In the pure Sullivan case, potentials $tw$ in $\mathbf{HH}^*(\mathcal{A},\mathcal{A})[[t]]$ correspond to odd-polyvector fields  

$$tw(d/de_1,d/de_2, \ldots d/de_m) \in T^{poly}(B)[[t]]$$

 The polyvector-field $v=\sum f(x_1,�,x_n)d/de_i$ in $T^{poly}(\bigwedge (V^*[1]))$ gives rise to a deformation of $(\bigwedge V^*[1])$. The general theory of deformations of Koszul dual algebras \cite{CalaqueII} shows that this deformation is isomorphic to $\mathcal{A}$. Since the Kontsevich formality map has the property that $$f_{n+1}(w,x_1,x_2,\ldots,x_n)=0, n>1 $$ this proves that $tw$ corresponds to the polyvector field we have claimed.  
 
After passing to the generic fiber, the Hochschild cohomology is given by :
$$(T^{poly}(V)((t)),[v+tw(d/de_1,\ldots,d/de_m),])$$

\begin{defn} By analogy with the case of ordinary matrix factorizations, we will say that $w$ has an \begin{em} isolated singularity \end{em} if the homology of this complex is finite dimensional. \end{defn}

Next we discuss the degeneration conjecture. For $(\mathcal{A},w)$ as above with isolated singularities, we can see from the formula for Hochschild cohomology that $HH^*(\mathcal{A},w)$ will always be concentrated in even degree. To see this, note that if we replace the variables $d/de_i$ by $u_i$ then we have a sequence of $\mathbb{Z}/2j\mathbb{Z}$ graded complexes 

$$ (T^{poly}(C[[u_1,u_2,\ldots]][x_1,\ldots,x_n]),[w+\sum f_iu_i,]) \subset HH^*(\mathcal{A},w)$$

  and  
  
     $$ HH^*(\mathcal{A},w) \subset (T^{poly}(C[x_1,\ldots,x_n][[u_1,u_2,\ldots u_m]]),[w+\sum f_iu_i,]) $$

In the special case when the cohomology of the complex is finite dimensional this implies that $w+\sum f_iu_i$ has isolated singularites. The degeneration conjecture is then automatic because the Hochschild cohomology is automatically concentrated in even degrees. As the category $MF(\mathcal{A},w)$ is Calabi-Yau, the Hochschild homology will all be concentrated in the either even or odd degree (depending upon the parity of the Calabi-Yau structure) and the degeneration conjecture thus follows for these algebras without any additional work. 

\begin{ex} For $\prod S^{2n_j}$ the condition that $w$ has an isolated singularity is similar to the usual Jacobian condition and states that $\mathbb{C}[u_1,\ldots, u_m]/(u_idw/du_i)$ be finite dimensional. The proof follows from the more general statement below. \end{ex}

More generally, we again discuss the case when $\mathfrak{g}$ is formal. In this case, recall that $\mathfrak{g}$ is determined by a bilinear form:

 $$B: \mathfrak{g}_{odd}\otimes \mathfrak{g}_{odd} \to \mathfrak{g}_{even}$$

and $U\mathfrak{g}$ is a graded Clifford algebra over $\mathbb{C}[u_1,\ldots,u_m]$. We let $D_k$ be the closed subvariety of $\mathbb{C}[u_1,\ldots u_m]$ for which $rank(B)\leq{k}$ and assume further that the $D_k-D_{k-1}$ is smooth. Let $R$ denote $U\mathfrak{g}/(w)$.

\begin{thm} Let $\mathcal{B}$ be a pure Sullivan algebra, whose Lie model $\mathfrak{g}$ is formal and as above. Let $w$ be a potential which intersects the varieties $D_k$ transversally at every point. Then: \begin{enumerate}[(a)] \item $w$ has isolated singularities 
                                                     
                                                       \item $Proj(R)$ has finite homological dimension as an abelian category.
                                                       \end{enumerate}  \end{thm}
The first statement is a calculation, so we explain the second one. Consider the exact functor between derived categories $$\pi: D^b(Gr-R) \to D^b(Proj(R)) $$

We can consider the abelian subcategory of $Gr-R$, denoted $Gr-R_{\geq i}$ which consists of modules $M$ such that $M_p=0$ for $p \leq i$ Restricted to this subcategory, 

$$\pi_{\geq i}: D^b_{\geq i}(Gr-R) \to D^b(Proj(R)) $$ 

has a right adjoint 

$$R\omega:D^b(Proj(R)) \to D^b_{\geq i}(Gr-R)$$

Thus we will show that for any $ M,N \in D^b(Gr-R)$, $\op{Ext}^i(M,R\omega \circ \pi(N))$ vanishes for large $i$. \vskip 10 pt 
 
Suppose that $Q$ is a graded prime ideal different from the maximal ideal and lying in a component of $D_k$, but not $D_{k-1}$. We denote $R/rad(QR)$ by $B$. Now denote by $P$ the prime ideal corresponding to the irreducible component of $D_k$ which $Q$ is in. One can prove that the correspondence $P \mapsto rad(PR)$ gives a bijection between (graded) prime ideals in $\mathbb{C}[u_1,\ldots,u_m]$ and (graded) prime ideals of $U\mathfrak{g}$ \cite{Musson}. We have a short exact sequence:

$$0\to S \to R/(rad(PR),Q) \to B \to 0$$ 

where $S$ is $B$ torsion by the assumption that the prime $Q$ lie in a component of $D_k$ but not $D_{k-1}$. Now we know by our condition, that $\mathbb{C}[u_1,\ldots,u_m]/Q[l]$ has a finite resolution as a $\mathbb{C}[u_1,\ldots,u_m]/P$ module and thus so does $R/(rad(PR),Q)$[l] as a $R/rad(PR)$ module. 

\vskip 10 pt The above exact sequence reveals that $\op{Ext}^i_{R/(rad(PR))}(B[l],M)$ is $B$ torsion for $i>m$. It is also easy to show from the transversality hypothesis that $R/rad(PR)[l]$ has finite homological dimension over $R$. Next, we note the following lemma, which is proved for ungraded rings in \cite{Brown} 

\begin{lem} Let $R$ be a graded FBN ring. Given a bounded complex C in $D(Gr-R)$ if $\op{Ext}^i(R/P[l],C)$ is $R/P$ torsion for $i>>j$ for every two-sided prime ideal $P$ then $\op{Ext}^i(M,C)$ vanishes for $i>>0$.\end{lem}

To finish the argument, we use a change of ring spectral sequence. Namely, we have a spectral sequence:

 $$ E^{pq} = Ext^p_{R/(rad(PR))}( R/rad(QR), Ext^q_{R}(R/rad(PR),M))  $$
 
 By the above discussion, $E^{pq}$ is $R/rad(QR)$ torsion for $p>m$. Because $R/rad(PR)$ has finite homological dimension over $R$, $E^{pq}$ vanishes for $q$ sufficiently high, depending only on $P$. Therefore for large enough $i$ only depending on $P$, $Ext^i_{R}(R/rad(QR),M)$ is torsion. Since there are only finitely many $P$ that arise, the result follows from the previous lemma.       


\section{Comments on the Pure Sullivan Condition}
The condition that our dg-algebra be pure Sullivan may seem like a restrictive condition. To get a better feeling for why this a natural condition if we are to expect a full open-closed field theory, we look at two examples, one where the rational homotopy type is hyperbolic and one where it is elliptic, but not pure Sullivan. 

 \begin{ex} Suppose now $\mathcal{Q}$ is $(S^3 \times S^3 \times S^3) \# (S^3\times S^3 \times S^3)$.
A standard calculation in rational homotopy theory proceeds as follows: \vskip 10 pt

Let $N$ be the wedge $(S^3 \times S^3 \times S^3) \vee (S^3\times S^3 \times S^3)$. \vskip 10 pt

Then it is clear that $$ \pi_*(\Omega N) \otimes \mathbb{Q} \cong Ab(x_1,x_2,x_3)* Ab(x_4,x_5,x_6) $$

In this formula, $Ab(x_i,x_j,x_k)$ denotes the abelian Lie algebra generated by three even variables and * denotes the free product of Lie algebras.
Next consider the manifold given by $ \mathcal{U} = S^3 \times S^3 \times S^3-D$, where $D$ is a small open disc in  $S^3 \times S^3 \times S^3$.

 $$ \pi_*(\Omega U) \otimes \mathbb{Q} \cong Ab(x_1,x_2,x_3)* Free(x) $$

Here $Free(x)$ denotes the free Lie algebra on one generator and $deg(x)=7$, which corresponds to the Whitehead triple product of three dimensional spheres.
Next we have the following general formula in \cite{Felix}, Theorem 24.7, for the rational homotopy Lie algebra of the connected sum of two manifolds $M$, $N$. 

$$ \pi_*(\Omega (M \# N)) \otimes \mathbb{Q} \cong \pi(\Omega M')*\pi(\Omega N')/(\alpha+\beta) $$ 

Here $M'$ and $N'$ are $M$ and $N$ with small discs removed and $\alpha$ and $\beta$ are the attaching maps for the top cell. \vskip 10 pt

In the case under consideration, the top cell is attached along the Whitehead product. This leads to the following calculation of homotopy groups for $\mathcal{Q}$:

$$ \pi_*(\Omega \mathcal{Q}) \otimes \mathbb{Q} \cong Ab(x_1,x_2,x_3)* Ab(x_4,x_5,x_6)* Free(x) $$

The center of the universal enveloping algebra can be seen to be $\mathbb{C}$ because the radical $R(\mathfrak{g})$ of the above Lie algebra is zero. 

\begin{lem} For any graded Lie algebra  $\mathfrak{g}$ in characteristic zero, such that each graded piece $ dim (\mathfrak{g}_i) < \infty $, there is a containment $ Z(U\mathfrak{g}) \subset U(R(\mathfrak{g}))$ \end{lem} 

Thus even on the homological level, the center of $H_*(\Omega \mathcal{Q})$ is given by $\mathbb{C}$. In view of the fact that the image of the map 

$$HH^*(C_*(\Omega \mathcal{Q})) \to H_*(\Omega \mathcal{Q})$$  

is contained in the center, there is no possibility for non-trivial curved deformations. 
\end{ex} 

 Even when the algebra is rationally elliptic and the image of the above morphism $HH^*(C_*(\Omega \mathcal{Q})) \to H_*(\Omega \mathcal{Q})$ is non-empty, there may be no compactifying deformation. Let $\mathfrak{g}$ be a nilpotent finite dimensional lie algebra concentrated in even degree. Let $\mathfrak{h}$ denote its center. 
 
 \begin{lem} Suppose the natural morphism $Sym(\mathfrak{h}) \to Z(U\mathfrak{g})$ is surjective. Then for any potential, $w$, the curved category $(\widehat{\mathcal{A}},w)-proj$ is either empty or non-compact. \end{lem}
 
 If $w$ has any linear component, then one can compute that the Hochschild cohomology vanishes. If $w$ is non-linear, let $I$ denote the ideal $\mathfrak{g}U\mathfrak{g} \cap Sym(\mathfrak{h})$. Next, consider the curved module $ M= \widehat{U\mathfrak{g}/I}$. We have that 
 
  $$ Hom(M,M) \cong \widehat{U\mathfrak{g}/I} \otimes \Lambda (h)  $$
  
To see this let $h_1,\ldots h_j$ denote a basis for $\mathfrak{h}$. We can write $w= \sum h_iw_i$.
 We let $(K_{(h_1,\ldots,j_j)}(\widehat{U\mathfrak{g}}),d_0)$ denote the Koszul complex associated to the ideal $I$.
Then the map $\wedge w_idh_i$ defines a map:

 $$ d_1: K^i_{(h_1,\ldots,j_j)}(\widehat{U\mathfrak{g}}) \to K^{i+1}_{(h_1,\ldots,j_j)}(\widehat{U\mathfrak{g}}) $$ 

One can then see that $d_1+d_0$ turns the $K_{(h_1,\ldots,j_j)}(\widehat{U\mathfrak{g}})$ into a matrix factorization $\mathcal{P}$.

$$ Hom(M,M) \cong Hom(\mathcal{P},M) \cong \widehat{U\mathfrak{g}/I} \otimes \Lambda (h)  $$   
 
We have $ dim_{\mathbb{C}}(Hom(M,M))= \infty$ and thus the category is not compact. \vskip 10 pt 
 
 Even if the map $Sym(\mathfrak{h}) \to Z(U\mathfrak{g})$ is not surjective, the above observation can be used to put strong restrictions on the possible curvings that can compactify the category. We do not pursue this further for reasons of space and interest.  \vskip 10 pt

\begin{ex} Let $\mathfrak{g}$ be a nilpotent finite dimensional lie algebra of rank three, with product given by $[x_1,x_2]=x_3$ and all other brackets are zero. The universal enveloping algebra is the algebra $\mathcal{A}=\mathbb{C} \lbrace x,y,z \rbrace /(xy-yx=z)$. Here the center of the universal enveloping algebra is a polynomial ring $\mathbb{C}[z]$ and we consider curved modules $(\mathcal{A},z^n)$. One can check that the category vanishes when $n=1$. 
One could also try to deform using not only curved deformation of $\mathcal{A}$ but also deform the higher multiplications.
 However, in this example, this does not affect the result.
  
 \begin{lem} There is no proper $\mathbb{Z}/2\mathbb{Z}$-graded deformation of $\mathcal{A}$. \end{lem}
 Any such Maurer-Cartan solution would necessarily be of the form 
 
 $$ p(z)+ p_{12}(x,y,z)d/dx\wedge d/dy + p_{13}(x,y,z)d/dx \wedge d/dz + p_{23}(x,y,z)d/dy \wedge d/dz $$ 
 
 where $p(z)$ is in $\mathbb{C}[z]((t))$ and $p_{ij}(x,y,z)$ are in $\mathbb{C}[x,y,z]((t))$. The fact that this satisfies the Maurer-Cartan equation implies that 
 
 $$p_{13}(x,y,z)= p_{23}(x,y,z)=0$$ 
  
 One can then compute that the  $p_{12}(x,y,z)d/dx\wedge d/dy$ terms are exact and conclude that there are no proper deformations. 
 \end{ex} 
 
 \section{Tangential Fukaya categories}

Given the close connection between string topology and the Floer theory of the cotangent bundle $T^*\mathcal{Q}$ explained in the introduction, we aim to give a Floer theoretic interpretation of our curved deformations of $C_*(\Omega \mathcal{Q})$. This construction was introduced independently by Nick Sheridan in his thesis \cite{She} and the author in \cite{Pomerleano}. \vskip 10 pt

For motivation, let us consider the easiest case of a symplectic mirror to a Landau-Ginzburg model, that of $S^2$. We think of a sphere as being the (open) disk bundle of the cotangent bundle, $D^*(S^1)$, compactified by the points at 0 and $\infty$. This is then mirror to 

$$(\mathbb{C}[z,z^{-1}], w= z+1/z)$$

Work of \cite{Seidel III} proves that if we want to understand mirror symmetry for the Landau-Ginzburg model 

$$(\mathbb{C}[z,z^{-1}], w= z^d+1/z^d)$$ 

we can either consider the Fukaya category of the orbifold $S^2//(\mathbb{Z}/d\mathbb{Z})$, where $\mathbb{Z}/d\mathbb{Z}$ acts by rotations that fix the two points, or more concretely a Fukaya category where we require disks to intersect the compactifying divisor with ramification of order $d$. \vskip 10 pt

This orbifold has a natural generalization. Consider a variety $X$ and a collection of effective Cartier divisor $\mathcal{D}_i$, and $d_i$ a collection of positive integers. The Cartier divisors define a natural morphism :

$$X\to [\mathbb{A}^n/(\mathbb{C}^*)^n]$$ 

\begin{defn} The \emph{root stack} $X_{(D_i,d_i)}$ is defined to be the fibre product 
 $$ X\times_{[\mathbb{A}^n/(\mathbb{C}^*)^n]}[\mathbb{A}^n/(\mathbb{C}^*)^n]$$ 
  
  where the map 
  
$$[\mathbb{A}^n/(\mathbb{C}^*)^n]\to [\mathbb{A}^n/(\mathbb{C}^*)^n]$$ is the $d_i$-power map. 

\end{defn} 
There are three important properties of the root stack:

\begin{enumerate}[(a)]

\item The root stack defines an orbifold, which has non-trivial orbifold stabilizers along the divisors 
\item The coarse moduli space is exactly $X$ and away from $D_i$ the map $X_{(\mathcal{D}_i,d_i)} \to X$ is an isomorphism. 
\item A map from a variety which is ramified to order $d_i$ along the divisors $\mathcal{D}_i$ lifts uniquely to a map to $X_{(\mathcal{D}_i,d_i)}$. \vskip 10 pt
\end{enumerate}

Let $\mathcal{Q}$ be any simply connected manifold with a metric whose geodesic flow is periodic. There are three known families of examples, that of $S^n$ $(n>1)$, $\mathbb{C}P^n$ and $\mathbb{H}P^n$. $T^*\mathcal{Q} - \mathcal{Q}$ then acquires a Hamiltonian $S^1$ action by rotating the geodesics (which then give rise to Reeb orbits when restricted to the unit cotangent bundle). This induces a natural Hamiltonian action on $(T^*\mathcal{Q}-\mathcal{Q}) \times \mathbb{C}$. The moment map for this Hamiltonian $S^1$ action 

 $$(T^*\mathcal{Q}-\mathcal{Q}) \times \mathbb{C} \to \mathbb{R}$$

is given by 
 
$$(x,z) \mapsto H(x)+1/2|z|^2$$ 

Where $H(x)=|x|$ is the Hamiltonian associated to the Hamiltonian action on $T^*\mathcal{Q} - \mathcal{Q}$. We then take the reduced space, that is the preimage of a regular value quotiented out by the $S^1$ action. Finally, we glue back in the zero section to obtain a manifold $X$ which is a symplectic compactification of the open disk bundle $D^*(\mathcal{Q})$ by the smooth divisor $\mathcal{D}$. \vskip 10 pt

When $\mathcal{Q}$ is $\mathbb{C}P^n$, $X \cong \mathbb{C}P^n \times \mathbb{C}P^n$. Namely, we have an anti-holomorphic involution, $$I:\mathbb{C}P^n\times \mathbb{C}P^n \to \mathbb{C}P^n \times \mathbb{C}P^n$$

 given by $$(z,w) \to (\bar{w},\bar{z})$$ 
 
 Its fixed point set: $ \mathcal{L} : \mathbb{C}P^n \to \mathbb{C}P^n \times \mathbb{C}P^n$, is a Lagrangian submanifold, which corresponds to the zero section in the general construction. The divisor $\mathcal{D}$ parameterizes oriented closed geodesics and is embedded as a $(1,1)$ hypersurface, the locus where 

$$\sum z_iw_i=0$$ 

When $\mathcal{Q}$ is $S^n$, we obtain the projective quadric $Q_n$ and the divisor $\mathcal{D}$ is the projective quadric $Q_{n-1}$. The zero section in the general construction corresponds to a vanishing sphere $\mathcal{L}$ under a degeneration to a singular quadric. \vskip 10 pt

Finally when $\mathcal{Q} \cong \mathbb{H}P^n$, we have that $X \cong Gr(2,2n+2)$, the Grassmanian variety of two-planes in $\mathbb{C}^{2n+2}$ \cite{Akhiezer}. \vskip 10 pt

In each of the cases we have $$ \pi_2(X,\mathcal{L}) \cong \mathbb{Z} $$

We want to consider three different moduli spaces of holomorphic disks.

\begin{defn} We define the moduli space of tangential disks to be $\mathfrak{M}^{d}_{j,\ell}(X,\mathcal{L})$ the moduli space of holomorphic maps whith the following extra data: 

           \begin{enumerate}[(a)] 
            \item a map $u: (\mathbb{D}^2,S^1) \to (X,\mathcal{L})$   
            \item  a collection of $j$ points on the boundary    
            \item  $u^{-1}(\mathcal{D})= d(p_1+ p_2+ \cdots p_{\ell})$
            where $p_j$ are points in $\mathbf{int}(\mathbb{D}^2)$
            \end{enumerate}
            \end{defn}
            
\begin{defn} We define the auxiliary moduli space $MA^d_{j}(X,\mathcal{L})$ which parameterizes objects which 
consist of the following three pieces of data: 
            
            \begin{enumerate}[(a)]
            \item  a collection of $j$ points $q_1, \ldots, q_j$ on the boundary of a disk
            \item two points $p_1$, $p_2$ in $\mathbf{int}(\mathbb{D}^2)$  such that there is a biholomorphism 
            
                       $$ \mathbb{D}^2 \to \mathbb{D}^2 $$ 
            
            which sends  $$ p_1 \to -r,\quad p_2 \to r,\quad q_1 \to i $$ 
            \item a map $(\mathbb{D}^2,S^1) \to (X,\mathcal{L})$ such that $u^{-1}(\mathcal{D}) = p_1+(d-1)p_2$
             \end{enumerate}
 \end{defn}
 
 \begin{defn} The Mickey Mouse moduli space $MM^{d}_{j,\ell}(X,\mathcal{L})$ parameterizes objects which 
consist of the following three pieces of data:   
  \begin{enumerate}[(a)]
  \item A map from a nodal disk $(u_1,u_2,u_3)$ with three components glued along marked points.
   \item a collection of marked points on the boundaries of each of the disks
  \item interior marked points $p_1$ and $p_3$ in $u_1$ and $u_3$ which intersect $\mathcal{D}$ with 
   multiplicities $d-1$ and 1. 
   \end{enumerate}
 \end{defn}
 
 See Sheridan's thesis \cite{She} for beautiful pictures of the above moduli spaces. \vskip 10 pt

 Using the moduli spaces of tangential disks, we now define a version of Floer theory for the Lagrangian submanifold $\mathcal{L} \subset X_{(\mathcal{D},d)}$. We take the point of view that we will count in our theory holomorphic disks which intersect the boundary divisor with multiplicity $d$. The difficulty in defining such a theory in general is that under Gromov compactness, holomorphic curves may have components consisting of holomorphic spheres which live entirely in the divisor $\mathcal{D}$ and the moduli space of such objects can often be non-regular since we cannot deform the complex structure in a neighborhood of the divisor.
In our situation this is not a problem because all moduli spaces of spheres in the divisor are automatically regular. 

\begin{lem} Let $\mathcal{D}$ be a compact complex homogeneous space, then all moduli spaces of nodal holomorphic spheres are automatically transverse.  \end{lem}

 In our situation, this problem is more manageable because such configurations have high codimension in the moduli space of $J$-holomorphic spheres. First, recall that a symplectic manifold $X$ is monotone if $\omega(X)= \tau c_1(TX)$ for $\tau \geq 0$. Because both $\mathcal{D}$ and $\mathcal{X}$ are monotone there is no problem with multiply covered curves. We have the following lemma proven by Ionel and Parker:

 \begin{lem} For moduli-spaces of somewhere injective spheres with prescribed intersection number with $\mathcal{D}$ it is possible to choose a complex structure which agrees with the above complex structure in a neighborhood of $\mathcal{D}$. \end{lem} 

In fact, as in the previous lemma it seems likely that one does not need to perturb the complex structure to acheive this, though we have not checked this. \vskip 10 pt  

Next, we have the following lemma: 
 
\begin{lem} The moduli space of simple tangential $J$-holomorphic spheres is generically a pseudo-manifold. The configurations of tangential $J$-holomorphic spheres consisting of non-constant components living inside the divisor is codimension at least 2. \end{lem}

We give the argument for $\mathcal{Q} = \mathbb{C}P^n$. The reader is encouraged to verify that the same argument is easily adapted to the case $\mathcal{Q}= S^n$ or $\mathbb{H}P^n$. The real dimension of the moduli-space of tangential $J$-holomorphic spheres with $l$ intersections with the divisor is:

  $$ 2n + 2c_1(u) - 6-2l(d-1)= 2n+2n(ld)-6+2l $$
  
We must analyze what happens to tangential spheres under Gromov compactness, Let $Y$ be the limit of a sequence of tangential spheres. Let $Z$ be a component contained containing marked points $z_1,\ldots, z_m$, and let $\alpha_i$ be the points in $Z$ which glue to components that intersect the divisor at isolated points with multiplicity $m_i$. $Y$ has the property that: 

$$ D\cdot Z + \Sigma_i m_i  = dm $$

 By a suitable perturbation, we can assume that the evaluation maps into the divisor $\mathcal{D}$ are transverse. Using the above dimension formula and the usual counting arguments for stable configurations of spheres, it is easy to verify that the codimension of such a configuration is $2 \cdot (k-1)$, where $k$ is the number of components of the tree. \vskip 10 pt

 As in relative Gromov-Witten theory, it is easy to acheive transversality for configurations of stable tangential disks, none of the components of which lie completely in the divisor. This is due to the following lemma in \cite{Cieliebak-Mohnke} which allows us to do all perturbations in a suitable open neighborhood $V$ of our Lagrangian $\mathcal{L}$.
 
 \begin{lem} There is a Baire dense set of tamed almost complex structures $\mathcal{J}^{reg}(V)$ that agree with $\mathcal{J}_0$ outside $V$ such that 
 $\mathfrak{M}^d_{j,\ell}(X,\mathcal{L})$  is regular.\end{lem}
 
It is interesting to note that in the case $\mathcal{Q}= \mathbb{C}P^1$, we have the following calculation: \vskip 10 pt

\begin{lem} $\mathfrak{M}^d_{j,\ell}(X,\mathcal{L})$ has the structure of an oriented pseudo manifold. \end{lem}

We make a transversality calculation for the open part of the moduli space in the case $\mathcal{Q}=\mathbb{C}P^1$. One can make a similar calculation for the various boundary components which arise under Gromov compactness. For this it is easier to work with the root stack $X_{(\mathcal{D},d)}$. 
We need to prove that the complex structure $J$ is regular for disks in $X_{(\mathcal{D},d)}$. We have a map from $\pi: X_{(\mathcal{D},d)} \to X$, which gives rise to an exact sequence $$ 0 \to TX_{(\mathcal{D},d)} \to \pi^*TX \to R \to 0 $$
For any map $f: \mathbb{D}^2 \to X_{(\mathcal{D},d)}$, we get a sequence of sheaves on $\mathbb{D}^2$, and using the reflection principle, we can double this to a sequence of sheaves on $\mathbb{C}P^1$. On $\mathbb{C}P^1$, we know that $R$ is a skyscraper sheaf of rank $2(d-1)\ell$, concentrated at the intersection point with $\mathcal{D}$ and its opposite and that that the  double of the  bundle $f^*TX$ is of the form $\mathcal{O}(2d\ell)\oplus \mathcal{O}(2d\ell)$. We conclude that the double of $TX_{(\mathcal{D},d)}$ is also positive to deduce the desired result. \vskip 10 pt

Again, the author does not know whether this persists for $S^{n+1}$ or $\mathbb{C}P^n$ for $n>1$, though as noted above this is somewhat tangential to our main line of inquiry.\vskip 10 pt 

In any case, the theory can be defined along standard lines in two equivalent ways, either following \cite{Seidel II} or \cite{FOOO}. \vskip 10 pt

If one were to follow the Morse-Bott definition in \cite{FOOO}, one would consider some model for chains, $C_*(X)((t))$, and using the evaluation maps (which for the purposes of discussion we assume to be transverse) 
$$ev_i: \mathfrak{M}^d_{k+1,\ell}(X,\mathcal{L}) \to \mathcal{L}$$ to define a sequence of higher products $$ m_k(\alpha_1,...\alpha_k)=\sum_{\ell} ev_{0,*}(\prod ev_i^*(\alpha_i))t^{\ell}$$

In our case, we are doing something slightly non-standard to our category by giving the Novikov-variable $t$ a grading in order to relate it to the deformations we considered previously. This is valid here because the Lagrangian $\mathcal{L}$ is a \begin{em} monotone \end{em} Lagrangian submanifold. Recall that a Lagrangian $\mathcal{L}$ is monotone if the two maps, corresponding to the action and Maslov index respectively:

$$ A: \pi_2(X,\mathcal{L}) \to \mathbb{R}, \text{      }     I: \pi_2(X,\mathcal{L}) \to \mathbb{Z}  $$  

satisfy the equation:

$$ 2A(u) =  \tau I(u) \text{        } \forall u \in \pi_2(X,\mathcal{L}) $$ 

In the above approach it is a somewhat technical issue to specify what types of chains one uses and how to define evaluation maps. It is therefore more convenient to use Seidel's approach of introducing inhomogeneous terms into the pseudoholomorphic curve equation. In our  \emph{Floer datum} \cite{Seidel II}, we will assume that  

\begin{itemize}
\item $H$ is a compactly supported Hamiltonian 
\item $J$ agrees with the above complex structure in a neighborhood of $\mathcal{D}$
\end{itemize}

Because the moduli space of J-holomorphic spheres in the divisor is already regular, we can achieve transversality with this type of Floer datum. In this regime, generators of the Floer complex correspond to time one chords of the Hamiltonian on $L$. We consider discs with n positive punctures $p_1, \cdots p_n$ and one negative puncture $q$. We attach striplike ends \cite{Seidel II} to our disc, that is a biholomorphisms $\mathbb{R}^+ \times [0,1]$ with a neighborhood of positive punctures and and a biholomorphisms $\mathbb{R}^- \times [0,1]$ with a neighborhood of the negative puncture. We also equip our disc with subclosed one forms $\beta$, which are equal to $dt$ in the striplike ends. We count zero dimensional moduli spaces of solutions to the perturbed equation. 

$$ (du-X_H \otimes \beta)_J^{0,1}=0 $$

Let $\gamma_i$ be generators of $CF(L,L,H_i)$. Then we define the $A_\infty$ product 

$$ m_n(o_{\gamma_1}  ) = \mathcal{M}(\gamma_1,\cdots, \gamma_n, \gamma_{n+1})o_{\gamma_{n+1}}  $$

Where $\mathcal{M}(\gamma_1,\cdots, \gamma_n, \gamma_{n+1})$ is the signed count of isolated solutions to the above equations which are asymptotic to $\gamma_i$ in the appropriate positive and negative striplike ends. Rules for determining orientations and signs again follow \cite{Seidel II}.\vskip 10 pt

With this summary behind us, using the perturbation lemma, we have defined an $A_{\infty}$ deformation:

$$ m_n: H^*(\mathcal{L})^{\otimes n} [[t]] \to H^*(\mathcal{L})[[t]] $$ 

the moduli spaces   $\mathfrak{M}^{d}_{j,1}(X,\mathcal{L})$ in particular define classes $e_d$ in $HH^*(C^*(\mathcal{Q}),C^*(\mathcal{Q}))$ 
 
\begin{lem} The class $e_d$ in $\mathfrak{M}^d_{j,1}(X,\mathcal{L})$ is gauge equivalent to the $d$-fold cup product $e_1^d$ of the class defined by $\mathfrak{M}^{1}_{j,1}(X,\mathcal{L})$ \end{lem}

To prove this result for all $d$, we proceed by induction and consider the auxiliary moduli space. By standard Gromov compactness arguments, the boundary of the auxiliary moduli space consists of points where: \vskip 10 pt 

 \begin{enumerate}
 [(a)] \item $r \to 0$, the boundary is the moduli space $\mathfrak{M}^{d}_{j,1}(X,\mathcal{L})$.
        \item $r \to 1$ the boundary is the Mickey Mouse moduli space $MM^d(X,\mathcal{L})$. 
  \end{enumerate} \vskip 10 pt
  
  The boundary as $r \to 1$ represents the Hochschild cup product, $e_{d-1} \cup e_1 $ \vskip 10 pt
 
 The boundary as $r\to 0$ is the class $e_d$. \vskip 10 pt
 
 This cobordism thus gives rise to the equation:
 
 $$ e_{d-1}*e_1 - e_d  = \partial(MA)  $$

where $MA$ is the Hochschild cochain defined by the auxillary moduli space. This equation implies the result by induction. \vskip 10 pt
  
It is interesting to note that the above proof follows the same line of reasoning as the proof in \cite{FOOO} that bulk deformation :
   
   $$ H^*(X) \to \mathbf{HH}^*(Fuk(X))$$
 is a ring homomorphism.\vskip 10 pt

When $d=1$ it is easy to calculate the relevant Hochschild cohomology class $e_1$.  In all cases, the cohomology ring is monogenerated. 
Namely we have that  $$ C^*(\mathcal{L}) \cong \mathbb{C}[x]/(x^j)$$

\begin{thm} The class $e_1$ is that which corresponds to the deformation $$ HF(\mathcal{L},\mathcal{L}) \cong \mathbb{C}((t))[x]/(x^j-t) $$ \end{thm} 

 For $\mathcal{Q}=\mathbb{C}P^n$ this follows from the fact that  $HF^*(\mathcal{L},\mathcal{L}) \cong QH^*(\mathbb{C}P^n)$, which is known to agree with the above ring \cite{FOOO}. For $\mathcal{Q}=S^n$, this result is contained in \cite{Smith}. We give a brief synopsis. To every \begin{em} weakly unobstructed \end{em} Lagrangian in $Fuk(X)$, we assign a number $m_0$ \cite{FOOO}. Let $Fuk(X,0)$ denote the subcategory generated by Lagrangians such that $m_0=0$. Smith shows that the Lagrangian $\mathcal{L}$ generates this subcategory. Let $QH^*(X,0)$ denote the 0-eigenspace under the map.
 
    $$ c_1 \cup: QH^*(X) \to QH^*(X) $$ 
 
  Smith proves that $QH^*(X,0) \cong HH^*(Fuk(X,0))$. It follows by Maslov index considerations that as a vector space $HF^*(\mathcal{L},\mathcal{L}) \cong H^*(S^n)$. Beauville \cite{Beauville} calculated the quantum cohomology of $X$. The only $A_{\infty}$ structure (up to gauge equivalence) on $H^*(S^n)$ which gives the correct $HH^*$ is the one above. \vskip 10 pt

Finally, we verify this if $\mathcal{Q}=\mathbb{H}P^n$. We have the following lemma for the Grassmannian $X=Gr(2,2n+2)$:

\begin{lem} $QH^*(X,0) \cong \mathbb{C}((t))[x]/(x^j-t)$ \end{lem}
To perform this calculation, we first note that there is the following presentation of the cohomology of the Grassmannian $X$.
Let $x_1$ and $x_2$ denote the Chern classes of the tautological bundle $E$ and $y_1,\ldots, y_{2n+2}$ denote the Chern classes of the complementary bundle $F$.
The fact that $E \oplus F$ is topologically trivial implies the following relation among the Chern classes for any positive integer $j$
 
 $$ \displaystyle \sum_i x_iy_{j-i} =0 $$
      
This relation combined with the vanishing of $y_{2n+1}$ and $y_{2n+2}$ gives the following presentation for the cohomology of the Grassmannian:

  $$ H^*(Gr(2,2n+2)) \cong \mathbb{C}[x_1,x_2]/(y_{2n+1},y_{2n+2})$$ 

 The quantum deformation is known to be given by $QH^*(X) \cong \mathbb{C}[x_1,x_2,t]/(y_{2n+1},y_{2n+2}-t)$ Focusing on the relation: 

 $$y_{2n+2} =x_2 \cdot y_{2n} + x_1 \cdot y_{2n+1} $$
 
 we observe that upon setting $x_1=0$, which corresponds to taking the 0-eigenspace of multiplication by the first Chern class, $y_{2n+2}=x_2^{n+1}$. This concludes the proof. \vskip 10 pt
 
 To finish the proof of the theorem we note that the gradings of $QH^*(X,0)$ and $HF^*(L,L)$ are well defined modulo $4n+4$ and the map $QH^*(X,0) \to HF^*(L,L)$ must preserve the gradings. Hence we conclude that this map is both injective and surjective since each element $1,x,x^2, \ldots,x^n$ in $QH^*(X,0)$ is invertible.  \vskip 10 pt
 
Returning to our main calculation, we have a ``finite determinacy" lemma. We state it for $\mathbb{C}P^n$, but the obvious adaptation of the theorem to the cases where $\mathcal{Q}= S^n$ or $\mathbb{H}P^n$ also holds. We first explain what we know about the $A(\infty)$ structure on $HF^*_{X_{D,d}}(\mathcal{L},\mathcal{L}) \cong \mathbb{C}[e]/e^{n+1}((t))$. For correctness, we note that for $d=1$ the above isomorphism is not an algebra map as the multiplication $m_2$ is deformed. Using the above lemma, we have the following formula for $e_d$

$$ e_d(e^{a_1},e^{a_2},\ldots,e^{a_{2d}})=t,\quad if \quad \sum (a_i)=(n+1)d$$

\begin{lem} The $A_{\infty}$ structure on $HF^*_{X_{D,d}}(\mathcal{L},\mathcal{L}) \cong \mathbb{C}[e]/e^{n+1}((t))$ is determined by the fact that $m_j=0$, $2<j<2d$ and $m_{2d}(e^{a_1},e^{a_2},\ldots,e^{a_{2d}})=t$, if $\sum (a_i)=(n+1)d$ \end{lem}
 
   We use the model for Hochschild cohomology developed in section 4 namely we notice that we have an equivalence Maurer-Cartan groupoids:

$$ MC^{\bullet} (HH^*(H^*(\mathbb{C}P^n))[[t]]) \cong  MC^{\bullet} (T^{poly}(C[x]\otimes \Lambda[\beta])[[t]], [x^{n+1} d/de,-]) $$
  
  Under this equivalence we have that the above deformation corresponds to a Maurer cartan element of the form  $$ T^{poly}(C[x]\otimes \Lambda[\beta],[x^{n+1} d/de,-])[[t]]$$ 
  
 of the form  
  
  $$ (d/de^d)t +\sum_k \tilde{m}^k t^k  $$
  
  For the lowest $k$ appearing in the sum above, it follows by a calculation of the Gerstenhaber algebra structure on $HH^*(C^*(\mathbb{C}P^n),C^*(\mathbb{C}P^n))$ that $[\tilde{m}^k, x^nd/de+(d/de^d t]=0$. \vskip 10 pt
  
 By calculations similar to those presented at the beginning of this section, $\tilde{m}^k$ is necessarily exact and thus there is a class $p$ such that $b(p)=\tilde{m}_k$. One can thus write down an $A(\infty)$ change of coordinates given by the formula $id+p$ which eliminates $\tilde{m}_k$ and one can see that this only effects products of higher degree. Continuing in this way, one can prove the desired lemma. \vskip 10 pt 

By a Kunneth theorem, we can get similar results for manifolds of the form $\mathcal{Q}=\prod \mathbb{C}P^{n_j}$. \vskip 10 pt

Finally, we explain how work in progress by Luis Diogo \cite{Diogo} gives a different approach to identifying the above defomations $e_d$. Such an approach is conceptually more satisfying, but is more analytically demanding than the approach we take above. We therefore explain these arguments without the necessary analytic detail, which is being developed in the above work. Given a monotone projective variety $X$ and a monotone ample divisor $\mathcal{D}$ as above with defining section $s$, we can consider the complement $\widehat{W}=X-\mathcal{D}$. The inverse image of a sufficiently large regular value $c$ of the exhausting function $f=-log||s||$ is a contact manifold and $W=f^{-1}((-\infty,c])$ is a Liouville domain. We give $\widehat{W}=X-\mathcal{D}$ the induced symplectic structure coming from the completion of this Liouville domain.  Let $b$ be a background class. Following Seidel, one can define symplectic cohomology of such a domain as well as open closed string maps: 

$$ \mathcal{CO}: SH^*(\widehat{W},b) \to HH^*(Fuk(\widehat{W},b))$$ 

 In the setting of a smooth ample divisor, one can arrange so that $\partial W$ is a circle bundle over $\mathcal{D}$ and the Reeb flow generates a circle action in a neighborhood of the divisor. Diogo considers J-shaped Hamiltonians $$H:\widehat{W} \to \mathbb{R}$$ such that \vskip 10 pt

\begin{enumerate}[(a)] \item on $W$, there is a value $r_1$ of the radial paramater $r$ defining the symplectic completion function such that $H$ is supported on
$(r_1,\infty) \times \partial W$.  
\item in the region $\partial W \times [r_1,\infty)$, $H(x,r)$ is a function $h(e^r)$, where $h$ is a strictly increasing function

\item $h''$ is positive, for $r> r_1$, and $lim_{r\to \infty} h'(e^r)=\infty$  \end{enumerate} \vskip 10 pt

Let $S(H)$ denote the set of time one periodic flows $\sigma$ of the Hamiltonian $H$. To make things explicit, in the region $\partial W \times [r_1,\infty)$, $X_H$ has the form $h'(e^r)R$, where $R$ is the Reeb vector field and time one orbits of $X_H$ can be identified with orbits of the Reeb vector field of time $h'(e^r)$. Thus, we have for every integer $d$, a Morse-Bott family of orbits $\partial W$ corresponding to $d$-fold iterates of the parameterized simple Reeb orbits as well as constant orbits corresponding to the interior of $W$. Using a Morse function $g$ on $\mathcal{D}$ and $\psi$ on $W$, we obtain a complex computing symplectic cohomology whose underlying vector space is given by:

$$ (CH \otimes_\mathbb{C} H^*(S^1) \oplus Morse(W,-\psi) [-n] , \operatorname{d}) $$

 \vskip 10 pt 

The vector space $CH$ is given by a copy of $Morse(\mathcal{D},g)$ for each positive natural number, which correspond to the space of Reeb orbits and their multiple covers. See Diogo's thesis for conventions concerning Morse homology. For simplicity, we assume that there is a perfect Morse function on $\mathcal{D}$ with a unique minimum, which we pullback to a generator of $H^0(\partial W)$. Diogo defines a complex which we call $SH^*_{alg}(\widehat{W},b)$, where the generators are the same as in symplectic cohomology for the J-shaped Hamiltonian, but the differentials are defined in terms of Gromov-Witten invariants of $\mathcal{D}$ and $X$. \vskip 10 pt

One important advantage of this model for our purposes is that for every multiplicity $d$, the chains which correspond to $H^0(\partial W)$  manifestly define \emph{classes} in $SH^*_{alg}(\widehat{W},b)$. After stretching the neck along the hypersurface, an SFT compactness result similar to the one in \cite{Bourgeois-Oancea} proves that there is a bijection of moduli spaces between the terms in the algebraic differential and the terms in the Floer cohomology differential for a sufficiently large value of the stretching parameter. \vskip 10 pt

We now consider the special case of the cotangent bundles considered in this section and explain how this gives a different approach to identifying our deformation classes. We can again stretch the neck to define a map:

$$ SH^*_{alg}(\widehat{W},b) \to HH^*(Fuk(\widehat{W},b))  $$

  An SFT compactness result similar to the one above should prove that this map agrees with the map $\mathcal{CO}$ for large values of the stretching parameter. The classes $e_d$ above are then the images of the class in $H^0(\partial W)$ for each multiplicity $d$. Namely, by a removal of singularities argument, punctured $\mathcal{J}$-holomorphic half cylinders with Lagrangian boundary asymptotic to $d$-fold covers of Reeb orbits in $T^*\mathcal{Q}$ are in bijection with with maps $\mathbb{D}^2\to X$ tangent to $\mathcal{D}$ with tangency of order $d$. It is easy to check that for $T^*S^n$, the computations in Diogo then agree with the results we obtained above. \vskip 10 pt

\begin{rem} In a closely related setup, $HH^*$ classes corresponding to non-trivial cohomology classes on $\partial W$ are used by Abouzaid and Smith in their study of Khovanov homology. \end{rem}

\section{Connections to Mirror Symmetry}

Our discussion in this section overlaps greatly with work of [Gross, Keel, and Hacking], which is motivated by tropical considerations. The analysis of the tangential Fukaya category suggests a mirror construction for a wide variety of Fano varieties, which we now explain. Let $X$ be a projective Fano variety and $D = \cup D_i$ an ample strict normal crossings divisor such that $c_1(X-D)=0$. We denote the complement $X-D$ by $U$.  We restrict to the simplest case where $D_i$ are individually ample and all of the strata $\cap D_I$ are monotone symplectic manifolds.  \vskip 10 pt
   
    To each component of the intersection, $D_{i_0} \cap D_{i_2} \cap \ldots \cap D_{i_p}$, we would like to define a Hochschild class of $Fuk(U)$, $\alpha_{([i_0,i_1,\ldots,i_p],j)}$, which counts holomorphic disks with Lagrangian boundary and which intersect the $j$-th component of the intersection $D_{i_0} \cap D_{i_2} \cap \ldots \cap D_{i_p}$ simply and intersect none of the other components. As we saw above, technical points need to be addressed to define these classes in general. A future paper will explain how to overcome these technical issues using ideas from Cieliebak and Mohnke's paper and discuss extensions of the classes to the wrapped Fukaya category $WFuk(U)$. In this section, we want to discuss the mirror symmetric interpretation of these classes.

 \begin{conj} Let $A$ be the subalgebra of $HH^*(WFuk(U))$ generated by the classes $\alpha_{I,j}$ above. For any object in the wrapped Fukaya category, $L$, the natural morphism $A \to HF(L,L)$ is module finite and $WFuk(U)$ can be given the structure of a Calabi-Yau category over $A$. \end{conj}

Let $X$ be a Fano variety and $[D]=[-K_X] \in Pic(X)$ , then one expects that the sub-algebra $A$ of Hochschild cohomology generated by the $\alpha_{I,j}$ coincides with global sections $\Gamma(\mathcal{O}_{U^{\vee}})$. In cases where $D$ has at least one intersection of maximal codimension, we can define the naive mirror of $U$, which we will denote by $U^{\vee}$, to be a crepant resolution of $Spec(A)$ should such a crepant resolution exist. 

\begin{rem} If we weaken the assumption that our divisor is monotone and work over the Novikov ring, this is consistent with a conjecture proposed in \cite{Auroux}. Namely, suppose that $D$ is a smooth anti-canonical divisor and the mirror, $U_q^{\vee}$, admits a proper map: 
$$\alpha: U_q^{\vee} \to \mathbb{A}_q^1$$  
the general fiber of which is mirror to $D$. The monomials of the form $\alpha^d$ inside of $\Gamma(O_{U_q}^{\vee})=\Lambda[\alpha]$ should correspond to the infinitesimal deformations underlying the tangential Fukaya category. 
\end{rem} \vskip 10 pt
 Technology which can be used to study the above conjecture has been developed by McLean\cite{McLean}, where he proves the following theorem. \begin{thm} Given a projective symplectic manifold $X$, and an ample normal crossings divisor $\cup D_i$, then $SH^*(U)$ has polynomial growth rate.   \end{thm} 

More precisely, the degree of growth is bounded by the maximal number of components which have non-empty intersection. As a special case of this result, we have that if $T^*\mathcal{Q}$ is symplectomorphic to an affine variety $U$, then $rank(H_k(L\mathcal{Q}))$ has finite growth rate. On the other hand, the author does not know of an example of such a cotangent bundle for which $\mathcal{Q}$ is not Pure Sullivan. One might therefore expect to derive additional restrictions on the symplectic topology of affine varieties. For example, a suitably strong version of the above conjecture would imply that the examples of section 6 could not arise as affine varieties. \vskip 10 pt
 
We now give a conjectural formula for the algebra $A$ and show that it reproduces several known mirror constructions. Our formula for the multiplication is again based upon consideration of the boundary of the auxillary moduli space will again yield the rules for multiplication. One component of the boundary will always correspond to the Hochschild multiplication. Unlike in the tangential Fukaya category above, sphere bubbles may arise which gives rise to an enumerative problem to compute the multiplication rules. \vskip 10 pt
 
 We have divisors $D_1,D_2, \ldots, D_j$. We consider tuples $(a_1,a_2, \cdots, a_j)$ which measures the intersection number of the class with the divisor $D_i$. If $I$ is a subset of $[1,2, \cdots, j]$, we consider the vector $(v_I)$ to be the vector such that $a_i=1$ if $i \subset I$ and zero otherwise. \vskip 10 pt

 If $I$ is a subset such that the intersection of $\cap_I D_i$ is non-empty, we assign a generator $\alpha_{I,m}$ to such a vector and a connected component of the intersection. We let $A$ be the free commutative ring generated by $\alpha_{I,m}$ with the following relations. Assume that either $I$ and $J$ are disjoint or that $D_I \cap D_J$ is empty. We then impose the relations that:

$$ \alpha_{I} \cdot \alpha_{J} = \sum_m \alpha_{I+J,m} + \sum_{K, B} GW_{D_K} (pt, D_I \cap D_K, D_J \cap D_K, [B]) \alpha_{K}   $$

Our formula needs several clarifications. The first is that if $D_I$ or $D_J$ is a collection of points, we have surpressed the components and our formula should be thought of as applying to each component individually. We also make a remark about how to interpret $\alpha_{I+J}$. Namely, if $D_I \cap D_J$ is empty, this is to be regarded as zero. We allow $K$ to be the empty set in which case we interpret $D_K$ as $X$ and $\alpha_{K}$ as 1. Finally, $B$ is the set of non-zero classes in $H_2(D_K)$ such that the vector 

$$ (B \cdot D_i) + v_K = v_I+v_J  $$

We record the following observation which will facilitate our computations. When $I+J$ above is a proper subset of $[1,2, \cdots, j]$, any holomorphic curve must intersect every component of the divisor.  We therefore conclude that the above formula specializes to :

$$ \alpha_{I} \cdot \alpha_{J}=\sum_m \alpha_{I+J,m} $$

 We now proceed to several examples.   \vskip 10 pt

Let $X = \mathbb{C}P^n $ and $D$ be the usual toric divisor. To each component of the divisor we assign generators $\alpha_i$. Then any possible holomorphic curve must necessarily intersect all of the toric divisors non-trivially. As a consequence, we have that the ring $A$ is defined by the single equation

$$ \prod_i \alpha_i=1 $$  \vskip 10 pt 

Next we consider the example where $X$ is $\mathbb{C}P^n$ and $D$ is the union of $n+2$ hyperplanes, $U$ is the generalized pair of pants spectacularly exploited by Sheridan in his proof of the homological mirror symmetry conjecture for hypersurfaces in projective space. To each divisor we again associate a variable $\alpha_i$. Then $$A \cong Jac(\mathbb{C}[\alpha_1, \alpha_2, \cdots, \alpha_{n+2}], \prod\alpha_i),$$ which is consistent with Sheridan's results.  \vskip 10 pt

Let $X$ be $\mathbb{C}P^n$ and let $D_1$ be a conic, while $D_2,\cdots, D_n$ are a collection of lines. From the formula above, we see that the only non-trivial relations are given by: 

$$ \alpha_1 \cdot \alpha_{[2,\cdots,n]} = \alpha_{[1,2,\cdots,n],0} + \alpha_{[1,2,\cdots,n],1} $$

$$\alpha_{[1,2,\cdots,n],0} \cdot \alpha_{[1,2,\cdots,n],1}= \alpha_{[2,\cdots,n]} $$ \vskip 10 pt

This latter equation is given by the holomorphic sphere $D_2 \cap D_3 \cdots \cap D_n.$ \vskip 10 pt

Moving closer to our examples, we consider $\mathcal{Q}^n$ and $D_1, \cdots, D_n$ are generic hyperplanes. The relations coming from Gromov-Witten theory of $X$ are given by 

$$ \alpha_1 \cdot \alpha_{[2,\cdots,n]} = \alpha_{[1,2,\cdots,n],0} + \alpha_{[1,2,\cdots,n],1} + 2   $$

$$ \alpha_{[1,2,\cdots,n],0} \cdot \alpha_{[1,2,\cdots,n],1} =1 $$ \vskip 10 pt

$Spec(A)$ is a singular variety, but is easily seen to have a crepant resolution $U^{\vee}$ as follows. $Spec(A)$ is isomorphic to the variety cut out by the equation: 

$$ \prod_{i=1}^n u_i = (1+w)^2 $$ 

where $w \subset \mathbb{C}^*$. This can be seen by replacing  $\alpha_{[1,2,\cdots,n],0}$ by $w$ and $\alpha_1=u_1/w$ and setting all other $\alpha_i=u_i$.  This is in turn the ring of invariants of an open substack of $(\mathbb{C}^n//\mathbb{Z}_2^{n-1})$ where $\mathbb{Z}_2^{n-1} \subset SL_n(\mathbb{C})$ acts diagonally. From the perspective of toric geometry, the fan of our singular affine toric variety $Spec(\mathbb{C}[u_1\cdots,u_n]^{\mathbb{Z}_2^{n-1}})$  can be thought of as a cone over a simplex where all of the sides are doubled. Choosing a unimodular subdivison gives rise to a crepant resolution $U^{\vee}$ of $Spec(A)$. \vskip 10 pt

From the above point of view, it seems reasonable to think of the Landau-Ginzburg model $$(U^{\vee}, u_2 + \cdots + u_n)$$ as the mirror to $T^*S^n$ and it is easy to see that we have an equivalence of categories. 

$$MF((U^{\vee}, u_2 + \cdots + u_n) \cong WFuk(T^*S^n)$$

Furthermore we expect an equivalence:

$$MF(U^{\vee}, u_1/w + u_2+ \cdots u_{n}) \cong Fuk(\mathcal{Q}^n) $$

 We now proceed to examine the consequences of this for our previous study. The reader will notice that the above results on the Fukaya category were not fully satisfactory as they described only the subcategory of $Fuk(\mathcal{Q}^n)$ generated by the zero-section $\mathcal{L}$ as a curved deformation, rather than the full Fukaya category. In this section, we show that by a slight modification of our construction, which roughly corresponds to remembering only the $\mathbb{Z}/2\mathbb{Z}$ grading on $C_*(\Omega S^{2})$, we may realize the full Fukaya category as a curved deformation. We then examine our construction using mirror symmetry to determine a possible geometric origin for our algebraic considerations. \vskip 10 pt 

 Recall that to any $\mathbb{Z}/2\mathbb{Z}$ graded curved algebra, Positselski has defined the absolute derived category of dg-modules over the curved algebra. Let $p_t(u)$ be a polynomial of the form $-ut + at^2u^3$ over the algebraic closure $K$ of $\mathbb{C}((t))$. Denote the critical values of $p_t(u)$ by $c_i$.  In the above notation we have that: 

\begin{thm} $$\prod_{\zeta=c_i,0} D^{abs}(K[e] ,p_t(u)-\zeta) \cong Fuk(\mathbb{C}P^1\times \mathbb{C}P^1,K) $$   \end{thm}

This is in fact obvious in view of our previous work and the fact that away from the origin the algebra $K[e]$ defines a trivial sheaf of Clifford algebras. This implies that at the critical values of $c_i$ the category is Morita equivalent to the category of modules over the ground field $K$. \vskip 10 pt

We consider two categories of matrix factorizations, which correspond to the $\mathbb{Z}$-graded and $\mathbb{Z}/2\mathbb{Z}$-graded modules over $C_*(\Omega S^{2})$ respectively. We realize $U^{\vee}$ as a blowup. 

$$U^{\vee} = \{ (u_1,u_2,w, t_1,t_2) \in \mathbb{C}^2 \times \mathbb{C}^* \times \mathbb{C}P^1 : u_1t_1=(1+w)t_2, u_2t_2=(1+w)t_1 \}$$

The critical locus of $u_2$ is contained within an affine subvariety where $t_2 \neq 0$. Therefore we can consider a new variable $q=t_2/t_1$. The neighborhood $t_1\neq 0$ is an affine space with variables $q,u$. In these coordinates, the potential $u_2=q^2u$. The object that we want to consider is the brane defined by $q=0$. Furthermore, a calculation reveals that 

$$ Hom(\mathbb{A}^1,\mathbb{A}^1) \cong  C_*(\Omega S^2)$$ 

Here is how that calculation goes. We can construct a matrix factorization \cite{Lin-Pomerleano} which resolves this coherent matrix factorization as follows:
\[P = ( \xymatrix{\mathbb{C}[u,q] \ar@/^/[r]^{q} & \mathbb{C}[u,q] \ar@/^/[l]^{uq} \cr } ) \]
From here we have that $Hom(P,P) \cong Hom(P,\mathbb{C}[u,q]/q)$. The differential vanishes on this latter complex and we have that
$$Hom(P,\mathbb{C}[u,q]/q) \cong \mathbb{C}[u] \otimes \Lambda(e)$$

Chasing through the isomorphism  $Hom(P,P) \cong Hom(P,\mathbb{C}[u,q]/q)$, we get that $m_{2}(e,e)=u$ as claimed. One point to remember is that we must view $C_*(\Omega S^2)$ as a $\mathbb{Z}/2\mathbb{Z}$ graded object. \vskip 10 pt

In the above notation, now just considering the category of matrix factorizations over $(\mathbb{C}[u,q],q^2u)$, it is also instructive to record the endomorphisms of the brane defined by $u=0$. The corresponding matrix factorization is given by:

\[P' = ( \xymatrix{\mathbb{C}[u,q] \ar@/^/[r]^{u} & \mathbb{C}[u,q] \ar@/^/[l]^{q^2} \cr } ) \]

Then we have that $Hom(P',P') \cong Hom(P',\mathbb{C}[u,q]/u) \cong \mathbb{C}[q]/q^2 $  \vskip 10 pt

This corresponds to the fact that the exceptional $\mathbb{C}P^1$ is mirror in $U^{\vee}$ to the zero section in $T^*S^2$.

Notice that there is $\mathbb{C}^*$ action on $U^{\vee}$, given by, 
$$\alpha \cdot (u_1,u_2,w)=(\alpha^{2} u_1,\alpha^{-2} u_2,w)$$ 
 
 Notice that $u_2$ has weight $-2$ with respect to this $\mathbb{C}^*$ action. We can use this $\mathbb{C}^*$- action to define a graded refinement of this category. Notice that with this grading, the variable $u_1$ has degree $2$, which makes the above isomorphism into a graded isomorphism. \vskip 10 pt

 Next we consider the Maurer-Cartan class $tu_1/w$. Notice that this is not a deformation of the category $MF_{\mathbb{C}^*}(U^{\vee},u_2)$ only of the category $MF(U^{\vee},u_2)$. We denote the $\mathbb{Z}/2\mathbb{Z}$ graded Hochschild cohomology by $HH^*_{\mathbb{Z}/2\mathbb{Z}}(C_*(\Omega S^2))$. \vskip 10 pt

Notice that the function $u/uq-1$ is different from $u$ even when restricted 
to the zero locus of $uq^2$. More precisely, we have that:

$$ u/(uq-1)= -u +u^2q \quad  mod(uq^2) $$

   Note that $HH_{\mathbb{Z}/2\mathbb{Z}}^*(C_*(\Omega S^2))$ is \emph{not} isomorphic to $HH_{\mathbb{Z}/2\mathbb{Z}}^*(C^*(S^2))$. To be precise, we can see from the analysis in section 1.2 that :
  
  $$ HH_{\mathbb{Z}/2\mathbb{Z}}^*(C_*(\Omega S^2))\cong \frac{\mathbb{C}[u,a]\otimes \Lambda b}{(a^2,ab,au)} $$

  $$ HH_{\mathbb{Z}/2\mathbb{Z}}^*(C^*(S^2)) \cong \frac{\mathbb{C}[[u]][a]\otimes \Lambda b}{(a^2,ab,au)} $$
  
The upshot of this discussion is that the classes $tu$ and $tu/w$ are equivalent as Hochschild classes in $HH^*_{\mathbb{Z}/2\mathbb{Z}}(C^*(S^2))[[t]]$, but not as Maurer-Cartan classes in $HH^*_{\mathbb{Z}/2\mathbb{Z}}(C_*(\Omega S^2))[[t]]$. We note that this claim does not contradict the finite determinacy lemma in the previous section. For both deformed categories mentioned above, $End(\mathbb{C}((t)),\mathbb{C}((t)))$ are the same. \vskip 10 pt

Geometrically, the equivariant category $MF_{\mathbb{C}^*}(U^{\vee},u_2)$ is mirror to the Wrapped Fukaya category consisting of exact Maslov-index zero Lagrangians. The mirror to the category $MF(U^{\vee},u_2)$ is given by considering a larger category which includes compact \begin{em} weakly unobstructed \end{em} Lagrangians, such as the Lagrangian tori used to define the Lagrangian torus fibration in \cite{AurouxII}. This category is now only $\mathbb{Z}/2\mathbb{Z}$-graded. \vskip 10 pt

 The cotangent fiber is still a generator for this larger category. Following \cite{AurouxII} there are exotic, non-exact Lagrangian tori contained $T^*S^2$ which are non-zero in the Fukaya category of $\mathbb{C}P^1\times \mathbb{C}P^1$ and which are Floer theoretically disjoint from the zero-section. These correspond to the other objects of the deformed category $MF(U^{\vee},u_2+ u_1/w)$ \vskip 10 pt

To formulate a precise theorem in this direction, we note the following lemma: 

\begin{lem} $-(u+u^2q)t \cong -ut+1/4u^3t^2$ as Maurer-Cartan classes in $HH^*(MF(\mathbb{C}[u,q,\frac{1}{uq-1}],uq^2)[[t]]$ \end{lem}

We apply the gauge transformation $-1/2t(u\partial_q)$  to the Maurer-Cartan class
$-t(u+u^2q)$. Recall that the formula for a gauge transformation is:
$$e^{-1/2tu\partial_q}(-(u+u^2q)t)= -t(u+u^2q) + \Sigma_{n\geq 0} \frac{1}{(n+1)!} ad^n(-1/2tu\partial_q)( [-1/2tu\partial_q,-(u+u^2q)t ]-[uq^2,-1/2tu\partial_q])$$ 
This simplifies to $p_t(u)=-ut+1/4u^3t^2$.  \vskip 10 pt
 
In view of the previous discussion it is interesting to discuss the root stacks $\mathbb{C}P^1\times \mathbb{C}P^1_{(D,d)}$ as well. For generic $t$, the LG-models ($U^{\vee},u_2+t(u_1/w)^d$) all have isolated critical locus. Let $Z$ denote the scheme theoretic critical locus of this function. \vskip 10 pt
 
 Recall that the orbifold cohomology $H^*(X_{(\mathcal{D},d)}\times_{(X_{(\mathcal{D},d)} \times X_{(\mathcal{D},d)})} X_{(\mathcal{D},d)})$ is the space of states for orbifold Gromov-Witten theory. Calculating the orbifold cohomology of $X_{(\mathcal{D},d)}$ we have the following lemma:

 \begin{lem} dim $ H_{orb}(\mathbb{C}P^1 \times \mathbb{C}P^1_{(D,d)}) = length(Z) $ \end{lem}
We assume $d>1$, since we have already worked out that case completely. It is easy to see that the rank of the left hand side is $2d+2$. For the right hand side one can again check that all of the critical points are contained in the chart with coordinates $u$ and $q$ as above.  Thus we must compute solutions to the equations:

$$ 2uq-td\frac{u^{d+1}}{(uq-1)^{d+1}} =0$$

$$q^2-td\frac{u^{d-1}}{(uq-1)^{d+1}}=0 $$

when $q=0,u=0$ the rank is $d+1$. Otherwise, we have that $uq=2$ 
and we have that $u^{d+1}= 4/td$.

 This observation should imply homological mirror symmetry for the orbifold once the appropriate definitions for the entire Fukaya category of $\mathbb{C}P^1 \times \mathbb{C}P^1_{(D,d)}$ are in place. \vskip 10 pt

 More precisely, the above lemma appears to be the shadow of a homomorphism:
 
  $$ H^*(X_{(\mathcal{D},d)}\times_{X_{(\mathcal{D},d)} \times X_{(\mathcal{D},d)}} X_{(\mathcal{D},d)}) \to HH^*(Fuk(X_{(\mathcal{D},d)}) $$
 
for some to be defined Fukaya category of the orbifold $X_{(\mathcal{D},d)}$.  In a different direction one expects these computations to generalize to higher dimensional $\mathcal{Q}^n$, though the author has not computed the non-zero modes $Fuk(\mathcal{Q}^n,c_i)$.\vskip 10 pt

  \section{Mirror Symmetry for the $A_n$ plumbings and curved deformations of plumbings}
  
 Previously, we have discussed the simplest sort of Stein manifold, cotangent bundles. In this section, we look at the next simplest case, which is the $A_n$ plumbing of cotangent bundles of $S^2$. Here the situation is considerably more complicated, and there is no known homotopical description of the wrapped Fukaya category. In this section, we compute the category for plumbings of spheres and then examine its curved deformation theory. \vskip 10 pt
 
 Again we consider the $A_n$ plumbing of spheres as a hypersurface: 

$$Y_\epsilon = \{ (x,y,z) \in \mathbb{C}^3 : xy-z^{n+1} = \epsilon\}$$ 

\begin{thm} $WFuk(Y_{\epsilon}) \cong MF_{\mathbb{C}^*}(\tilde{X},v)$ \end{thm}

 The space $(\tilde{X},v)$ is the mirror of $Y_{\epsilon}$, whose construction follows closely the ideas of the previous section. More precisely, the map

 $$ z: Y_{\epsilon} \to \mathbb{C} $$

defines a Lefschetz fibration.

One can then construct a special Lagrangian fibration on $Y_\epsilon$ to obtain the mirror manifold. For a detailed exposition of how to construct the mirror using this special Lagrangian fibration, the reader may consult \cite{Chan}, based upon an earlier talk of Auroux \cite{Auroux II}. \vskip 10 pt

 We again consider the minimal resolution $\tilde{X}$ of the $A_n$ singularity,

 $$X = \{ (u,v,w) \in \mathbb{C}^2 \times \mathbb{C}^* : uv-(1+w)^{n+1} = 0\}$$
 
 The mirror is best described as an open part of a toric variety. Let

 $$X_{tor} = \{ (u,v,w) \in \mathbb{C}^2 \times \mathbb{C} : uv-(1+w)^{n+1} = 0\}$$
 
The fan for this toric singularity is given by  $\varDelta$ in $N_\mathbb R$ for $N=\mathbb Z^2$ consisting of the cone generated by the vectors:\\
\[	v_{\rho_1}=(0,1),\; v_{\rho_2}=(n,1),	\]\\

The toric resolution is given by the fan $\varDelta$ in $N_\mathbb R$ for $N=\mathbb Z^2$ consisting of all cones generated by no more than two of the vectors:\\
\[	v_{\rho_1}=(0,1),\; v_{\rho_2}=(1,1),\; v_{\rho_3}=(2,1),\; \ldots v_{\rho_{n+1}}=(n,1)		\]\\

The mirror is equipped with a natural $\mathbb{C}^*$ action:

$$\alpha * (u,v,w)=(\alpha^{2} u,\alpha^{-2} v,w)$$ 
 
 It is with respect to this action that we consider the category of graded D-Branes. The zero fiber of $v$ is an $A_n$ configuration of $\mathbb{C}P^1$ in the fiber of the resolution over $(0,0,-1)$ in $\tilde{X} \to X$, the components of which we denote by $Z_i$, $i=1,\ldots, n$ glued to an $\mathbb{A}^1$, which we denote by $A$ at a single point along $Z_1$. The reduced scheme structure on the singular locus of $v$ coincides with $Z_i \cup A$ for $i=1,\ldots, n-1$. Under the mirror map, each $\mathbb{C}P^1$ in the zero fiber corresponds to a matching sphere Lagrangian and the $\mathbb{A}^1$ corresponds to a Lefschetz thimble at the beginning of the chain. \vskip 10 pt

  We will need to study the Fukaya-Seidel category $Fuk(Y_{\epsilon},z)$. We recall its definition here. Objects consist of (twisted complexes of) exact Lagrangians in $Y_{\epsilon}$ and an ordered collection of Lefschetz thimbles $\Delta_1$, $\Delta_2$, $\ldots \Delta_{n+1}$. Let $ V_1 $, $V_2$, $\ldots V_{n+1}$ denote the corresponding vanishing cycles. In his book \cite{Seidel II}, Seidel defines the Floer cohomology for the thimbles as: 
  
    \[
Hom(\Delta_i,\Delta_j)=\left\{\begin{array}{cl}
\mathbb{C},& i=j \\
HF^*(V_i,V_j), & i<j \\
0,             & i>j      
\end{array} \right.\] 
  
  \vskip 10 pt
  
We apply the following theorem of Abouzaid and Seidel, which enables us to compute the wrapped Fukaya category of the Lefschetz fibration in terms of the inclusion of a smooth fibre and the Fukaya-Seidel category $Fuk(Y_{\epsilon},z)$ \cite{Abouzaid2, Seidel4} :  

\begin{thm} There is a natural transformation from Serre: $[Fuk(Y_{\epsilon},z)] \to [Fuk(Y_{\epsilon},z)]$ to the $id: [Fuk(Y_{\epsilon},z)] \to [Fuk(Y_{\epsilon},z)]$ such that $WFuk(Y_{\epsilon})$ is isomorphic to the localization of of $Fuk(Y_{\epsilon},z)$ with respect to this natural transformation. \end{thm} 

We describe how this works. We denote by $B$ the exceptional algebra associated to the thimbles in $Fuk(Y_{\epsilon},z)$. Let $E$ be the algebra associated to the vanishing cycles in the fiber $Fuk(Y_z)$. We have an inclusion $B \to E$ of $A_\infty$ algebras. We have an exact sequence of $B-B$ bimodules 

                  $$B \to E \to E/B $$

taking the boundary morphism $E/B \to B$ gives rise to the above natural transformation. \vskip 10 pt

 The strategy we will follow is to first produce a mirror to $[Fuk(Y_{\epsilon},z)]$ and then use the above theorem of Seidel and Abouzaid to deduce the result. We prove the result by partially compactifying $\tilde X$ by adding an extra divisor $D$, which corresponds to adding the vector $(-1,0)$ to the above toric fan. 

This is mirror to the Lefschetz fibration $(Y_{\epsilon},z)$. Notice that the divisor $D$ is isomorphic to $\mathbb{C}^*$ which is mirror to the fibre $Y_z \to Y_{\epsilon}$ (also a $\mathbb{C}^*$). 
\vskip 10 pt

We denote this compactified space by $\overline{X}$. Notice that the divisor $D$ is an anti-canonical divisor for $\overline{X}$. The singular locus of the potential $v$ when extended to the space $\overline{X}$ is a chain of $\mathbb{C}P^1$, whose irreducible components consist of the compactification of $A$ denoted by $\bar{A}$ and $Z_i$. \vskip 10 pt

Let $L$ denote $\mathcal{O}_{\bar{A}} \oplus_i \mathcal{O}_{Z_i}$, $i=1,\ldots n$ as objects in $MF(\overline{X},v)$. The object $L$ generates the category because it is proven in \cite{Lin-Pomerleano} that a generator of the category of coherent sheaves on the singular locus generates the category $MF(\overline{X},v)$. Picking a point $p$ away from the singular locus on $Z_n$, we have that the collection $p \oplus \mathcal{O}_{Z_n}$ generates $Coh(Z_n)$. In particular, it generates the skyscraper sheaf at the point $q= Z_n \cap Z_{n-1}$. Inductively, we can now generate the category $Coh(\cup Z_i)$, where $i=1,\ldots n-1$. The object $[p]$ is zero in $MF(\overline{X},v)$, since it avoids the critical locus. \vskip 10 pt

It is interesting to notice that in the case of $MF(\tilde{X},v)$, the category of matrix factorizations is actually generated by $\mathcal{O}_{A} \oplus_i \mathcal{O}_{Z_i}$ $i=1,\ldots n-1$. The reason is that $A$ is affine so we can generate the skycraper sheaf at $q'= A \cap Z_1$. Again using an inductive argument, we can generate the category of coherent sheaves on the entire critical locus. \vskip 10 pt   

Similarly, let $\mathcal{L}$ denote the sum of $\Delta_1$ and the matching spheres $\mathcal{L}_j$. Seidel proves that the Lefschetz thimbles split generate the Fukaya-Seidel category \cite{Seidel II}. As noted above, in our setting it is more appropriate to consider a different generating set for the Fukaya category. The thimble $\Delta_1$ and the matching spheres $\mathcal{L}_j$ also generate the Fukaya-Seidel category $Fuk(Y_{\epsilon},z)$. This can be seen because given a thimble at the beginning of the chain and all of the matching spheres, the other thimbles arise as a Dehn twist \cite{Seidel II} of the preceeding thimble and the matching sphere. \vskip 10 pt

 It is easy to verify on the homology level that: $$ Hom_{MF(\overline{X},v)}(L,L) \cong Hom_{Fuk(Y_\epsilon,z)}(\mathcal{L},\mathcal{L}) $$ The homology $Hom_{MF(\overline{X},v)}(L,L)$ has the following quiver presentation. 


\[Q = ( \xymatrix{\alpha_0 \ar@/^/[r] & \alpha_1 \ar@/^/[l] \ar@/^/[r] & \alpha_2 \ar@/^/[l] \ar@/^/[r] &\cdots \ar@/^/[l] \ar@/^/[r] & \alpha_{n-1} \ar@/^/[r] \ar@/^/[l] & \alpha_n \ar@/^/[l]} )\]


The quiver is given by taking the path ring of the above graph modulo the following relations

$$ (\alpha_0|\alpha_1|\alpha_0)=0, (\alpha_i|\alpha_{i+1}|\alpha_i)= (\alpha_i|\alpha_{i-1}|\alpha_i)$$ 
  
   and 

$$ (\alpha_{i-1}|\alpha_i|\alpha_{i+1})= (\alpha_{i+1}|\alpha_{i}|\alpha_{i-1})=0 $$

Our grading conventions follow those of the foundational \cite{Seidel-Thomas}. We next explain how to verify this. The calculation of $Hom_{MF(\overline{X},v)}(\mathcal{O}_{Z_i},\mathcal{O}_{Z_i})$ proceeds by analogy with the corresponding calculation in the ordinary derived category. As was proven in \cite{Lin-Pomerleano}, we can compute $Hom(\mathcal{O}_{Z_i},\mathcal{O}_{Z_i})$ by $R\Gamma(R\mathcal{H}om(\mathcal{O}_{Z_i},\mathcal{O}_{Z_i}))$. The local calculation in the previous section demonstrates that $$ R\mathcal{H}om(\mathcal{O}_{Z_i},\mathcal{O}_{Z_i}) \cong \mathcal{O}_{Z_i} \oplus \mathcal{N}_{Z_i}$$. Thus $Hom^0(\mathcal{O}_{Z_i},\mathcal{O}_{Z_i}) \cong \mathbb{C}$. \vskip 10 pt

 The fact that the branes $\mathcal{O}_{Z_i}$ do not intersect the compactifying divisor implies that: 
 
 $$Hom^2(\mathcal{O}_{Z_i},\mathcal{O}_{Z_i}) \cong \mathbb{C}$$
 
  For $\mathcal{O}_L$, $\mathcal{N}_{L} \cong \mathcal{O}(-1)$ and $Hom^2(\mathcal{O}_{L},\mathcal{O}_{L})$ vanishes. The calculations that 
  
  $$Hom_{MF(\overline{X},v)}(\mathcal{O}_{Z_i},\mathcal{O}_{Z_{i+1}}) \cong \mathbb{C}$$
  
    and 
 
 $$Hom_{MF(\overline{X},v)}(\mathcal{O}_{Z_{i+1}},\mathcal{O}_{Z_{i}}) \cong \mathbb{C}$$
 
 are also purely local. If the natural equivariant structure on each of these objects is taken into account, the gradings also agree with those listed above. \vskip 10 pt
 
We also have the following formality lemma for the $Hom_{MF(\overline{X},v)}(L,L)$ in the matrix factorization category, whose proof is an adaptation of an argument of Seidel and Thomas \cite{Seidel-Thomas}:  
\begin{lem} The quiver algebra $Hom_{MF(\overline{X},v)}(L,L)$ is intrinsically formal. \end{lem}

We summarize the main point. Lemma 4.21 of the paper by Seidel and Thomas prove that the $A_{n+1}$ quiver algebra, which we denote by $A$ is formal. We denote the quiver that we care about by $\tilde{A}$. For degree reasons we have the exact sequence:
 
 $$ \mathbf{HH}^{q-1}(A,A[2-q]) \to \mathbf{HH}^{q}(A,A[2-q]) \to 0 $$

 Using their notation, we define $\phi_{i_0,i_1,\ldots,i_0} \in \mathbf{HH}^q(A,A[2-q])$ to be 
 
    \[
\phi_{i_0,i_1,\ldots,i_0}(c) =\left\{\begin{array}{cl}
(i_0|i_0+1|i_0),& c= (i_0|i_1)\otimes \ldots \otimes (i_{q-1}|i_0) \\
0 , & \text{all other basis elements c}\\
\end{array} \right.\] 
  
  \vskip 10 pt

These form a basis for $\mathbf{HH}^q(A,A[2-q])$. Seidel and Thomas use two sets of classes of Hochschild cochains in $\mathbf{HH}^{q-1}(A,A[2-q])$ to generate enough relations in $HH^q(A,A[2-q])$ to prove that this group is zero. These classes are: 
    
 \[
\phi'(c) =\left\{\begin{array}{cl}
(i_0|i_{q-2}),& c= (i_0|i_1|)\otimes (i_2|i_3)\ldots \otimes (i_{q-3}|i_{q-2}) \\
0 , & \text{all other basis elements c}\\
\end{array} \right.\]   
    
     \[
\phi''(c) =\left\{\begin{array}{cl}
(i_0|i_1|i_0),& c= (i_0|i_1|i_0)\otimes (i_3|i_4)\ldots \otimes (i_{q-2}|i_0) \\
0 , & \text{all other basis elements c}\\
\end{array} \right.\] 

\vskip 10 pt

 The first set of classes are uneffected by annihilating $(\alpha_0|\alpha_1|\alpha_0)$. The second set of classes are precisely those classes which annihilate $\phi_{\alpha_0,\alpha_1,\ldots,\alpha_0}$, which also vanish when we annihilate $(\alpha_0|\alpha_1|\alpha_0)$. Thus, the same argument implies that $HH^2(\tilde{A},\tilde{A})=0$.

\vskip 10 pt

It follows that after idempotent completion, $Fuk(Y_\epsilon,z) \cong MF(\overline{X},v)$. From here, we note that the divisor $D$ defines a section of the anti-canonical line bundle $\mathcal{K}^{-1}$ on $\overline{X}$ and hence a natural transformation. This natural transformation is given by taking a matrix factorization $P$, tensoring it with $\mathcal{K}$, and using the section defining our divisor $\mathcal{D}$ to define a morphism:

   $$P\otimes \mathcal{K} \to P $$
   
 \begin{lem} $MF(\tilde{X},v)$ is the localization of $MF(\overline{X},v)$ with respect to this natural transformation. This natural transformation is mirror to that induced by the inclusion of the fibre $Y_z \subset Y_\epsilon$.\end{lem}
 To begin, we observe that the functor 
  
  $$ \pi: MF(\overline{X},v) \to MF(\tilde{X},v) $$
 
 is essentially surjective because coherent sheaves extend from open subsets. Next, let $\mathcal{C}$ be a category, $F$ a functor and $N:F \to id$ a natural transformation. 
 Let $\pi$ denote the functor 
 
 $$ \mathcal{C} \to \mathcal{C}_{loc} $$ 

where $\mathcal{C}_{loc}$ denotes the localization with respect to $N$. Then we have that:   

 $$ \varinjlim Hom_{\mathcal{C}}(F^p(X),Y) \cong Hom_{\mathcal{C}_{loc}}(\pi(X),\pi(Y)) $$
  
Recall that given two matrix factorizations, we can compute $Hom(E,F)$ by $R\Gamma(\mathcal{H}om(E,F))$. The result now follows since:
 
 $$ R\Gamma(\mathcal{H}om(E,F)|_{\tilde{X}}) \cong \varinjlim R\Gamma(\mathcal{H}om(E,F\otimes K^{-p}))$$ 

 To show that the natural trasformations are equivalent, note that the natural transformations are trivial restricted to the $\mathcal{O}_{Z_i}$.
Thus, we only have to consider the object $\mathcal{O}_{\bar{A}}$. By the matrix factorization calculations mentioned above, we have $Hom^0(\mathcal{O}_{\bar{A}},\mathcal{O}_{\bar{A}}(1)) \cong \Gamma(\mathcal{O}_{\bar{A}}(1))_0 \cong \mathbb{C}$. Here $\Gamma(\mathcal{O}_{\bar{A}}(1))_0$ denotes the invariant part of $\mathcal{O}_{\bar{A}}(1)$ with respect to the natural induced equivariant structure on $\mathcal{O}_{\bar{A}}(1)$. From this information it is easy to conclude that the two natural transformations give rise to isomorphic categories after localization.  \vskip 10 pt

To demonstrate the non-trivial nature of the wrapped Fukaya category of the plumbing, one can compute directly the endomorphisms of the Lefschetz thimble in the wrapped category. For example, the thimble at the beginning of the chain discussed above has endomorphism algebra:

 $$\mathbb{C}[u] \otimes \Lambda(e) \quad m_{n+1}(e,e,e...,e)= u^n$$
 
We conclude our thesis by stating the following proposition, which the reader can check by direct computation.

\begin{lem} The curved deformations corresponding to $u^j$ compactify the category. \end{lem}


\end{document}